\documentclass[aps,prl,english,amsmath,amsfonts,amssymb,superscriptaddress,twocolumn,showpacs,citeautoscript,reprint,longbibliography]{revtex4-1}
\usepackage[T1]{fontenc}
\usepackage[latin9]{inputenc}
\usepackage{amsmath}
\usepackage{amssymb}
\usepackage{wasysym}
\usepackage{adjustbox}
\usepackage{esint}
\usepackage{graphicx}
\usepackage{times}
\usepackage{nicefrac}
\usepackage{appendix}
\usepackage{multirow}
\usepackage{textcase}
\usepackage{subfigure}
\usepackage{empheq}
\usepackage{color}
\usepackage{leftidx}
\usepackage{hhline}
\usepackage{makecell}
\usepackage{stackrel}
\usepackage{lineno}
\makeatletter
\@ifundefined{textcolor}{}
{%
 \definecolor{BLACK}{gray}{0}
 \definecolor{WHITE}{gray}{1}
 \definecolor{RED}{rgb}{1,0,0}
 \definecolor{GREEN}{rgb}{0,1,0}
 \definecolor{BLUE}{rgb}{0,0,1}
 \definecolor{CYAN}{cmyk}{1,0,0,0}
 \definecolor{MAGENTA}{cmyk}{0,1,0,0}
 \definecolor{YELLOW}{cmyk}{0,0,1,0}
}

\renewcommand{\vec}[1]{\mathbf{#1}}
\renewcommand{\Re}{\operatorname{Re}}
\renewcommand{\Im}{\operatorname{Im}}
\newcommand{\tr}{\operatorname{Tr}}

\newcommand{\asym}{\operatorname{Asym}}

\renewcommand{\b}{\beta}

\newcommand{\add}[1]{\if\a\b{{\color{red} #1}}\else{#1}\fi}

\newcommand{\bracket}[1]{\langle #1 \rangle}
\newcommand{\ket}[1]{| #1 \rangle}
\newcommand{\bra}[1]{\langle #1 |}
\newcommand{\im}{\mathrm{i}}

\renewcommand{\eqref}[1]{(\ref{eq:#1})}

\newcommand{\figref}[1]{Fig.~\ref{fig:#1}}

\newcommand{\TT}{\mathbb{T}}
\newcommand{\Ss}{\mathbb{S}}
\newcommand{\VV}{\mathbb{V}}
\newcommand{\GG}{\mathbb{G}}

\newcommand{\YY}{\mathbb{Y}}
\newcommand{\II}{\mathbb{I}}

\newcommand{\Gvac}{\mathbb{G}^{\mathrm{vac}}}

\makeatother
\usepackage{babel}

\begin{document}

\title{Fundamental limits to radiative heat transfer: theory}

\author{Sean Molesky} \thanks{These two authors contributed equally.}
\author{Prashanth S. Venkataram} \thanks{These two authors contributed equally.}
\author{Weiliang Jin}
\author{Alejandro W. Rodriguez}
\affiliation{Department of Electrical Engineering, Princeton
  University, Princeton, New Jersey 08544, USA}

\date{\today}

\begin{abstract}
  Near-field radiative heat transfer between bodies at the nanoscale
  can surpass blackbody limits on thermal radiation by orders of
  magnitude due to contributions from evanescent electromagnetic
  fields, which carry no energy to the far-field. Thus far, principles
  guiding explorations of larger heat transfer beyond planar
  structures have assumed utility in surface nanostructuring, which
  can enhance the density of states, and further assumed that such
  design paradigms can approach Landauer limits, in analogy to
  conduction. We derive fundamental shape-independent limits to
  radiative heat transfer, applicable in near- through far-field
  regimes, that incorporate material and geometric constraints such as
  intrinsic dissipation and finite object sizes, and show that these
  preclude reaching the Landauer limits in all but a few restrictive
  scenarios. Additionally, we show that the interplay of material
  response and electromagnetic scattering among proximate bodies means
  that bodies which maximize radiative heat transfer actually maximize
  \emph{scattering} rather than absorption. Finally, we compare our
  new bounds to existing Landauer limits, as well as limits involving
  bodies maximizing far-field absorption, and show that these lead to
  overly optimistic predictions. Our results have ramifications for
  the ultimate performance of thermophotovoltaics and nanoscale
  cooling, as well as related incandescent and luminescent devices.
\end{abstract}

\maketitle 

The concept of a blackbody, derived from electromagnetic reciprocity
(or detailed balance), has provided a benchmark of the largest
emission rates that can be achieved by a heated object: through
nanoscale texturing, gray objects can be designed in myriad ways to
mimic the response of a blackbody at selective
wavelengths~\cite{JinPRB2019, ConstantiniPRAPP2015, RodriguezPRL2011},
with implications for a variety of technologies, including
high-efficiency solar cells, selective emitters, and thermal
sensors~\cite{AsanoSCIADV2016}. Over the past few decades, much effort
has gone toward understanding analogous limits to enhancements of
near-field radiative heat transfer (RHT)~\cite{PendryJPCM1999,
  BiehsPRL2010, BenAbdallahPRB2010, MillerPRL2015}, supported by a
rich and growing number of experimental~\cite{ShenNANOLETT2009,
  StGelaisNANOLETT2014, CuiNATURE2017, KloppstechNATURE2017} and
theoretical~\cite{LuoPRL2004, OteyJQSRT2014, PolimeridisPRB2015,
  RodriguezPRB2013, KrugerPRB2012} investigations, and motivated by
potential applications to
thermophotovoltaics~\cite{LenertNATURENANO2014, KaralisSR2016},
nanoscale cooling~\cite{GuhaNANOLETT2012}, and thermal
microscopy~\cite{BoudreauRSI1997, JonesNANOLETT2012}. A key principle
underlying further near-field RHT enhancements is the use of materials
supporting bound (plasmon and phonon) polaritons in the infrared,
where the Planck distribution peaks at typical temperatures probed in
experiments. This leads to strong subwavelength responses tied to
corresponding enhancements in the density of
states~\cite{VolokitinPRB2001, DominguesPRL2005, VolokitinRMP2007,
  SongNATURENANO2015}; consequently, the amplified near-field RHT
spectrum exhibits a narrow lineshape, justifying focus on selective
wavelengths. However, while the properties of such polaritons,
particularly their resonance frequencies, associated densities of
states, and scattering characteristics can be modified through
nanoscale texturing, only recently have computational
methods~\cite{ReidPROCIEEE2013, OteyJQSRT2014, RodriguezPRB2013,
  PolimeridisPRB2015} arisen to model RHT between bodies of arbitrary
shapes beyond those with high symmetry~\cite{KrugerPRB2012,
  PerezMadridPRB2008, MessinaPRB2013}. Furthermore, the challenge of
gaining simultaneous control over the scattering properties of large
numbers of contributing surface waves has generally precluded general
upper bounds on RHT.

RHT between two bodies A and B in vacuum is given as
\begin{equation} \label{eq:integratedRHT}
  P = \int_{0}^{\infty}
  [\Pi(\omega, T_{\mathrm{B}}) - \Pi(\omega, T_{\mathrm{A}})]
  \Phi(\omega)~\mathrm{d}\omega,
\end{equation}
in terms of their local temperatures $T_{\mathrm{A}}$ and
$T_{\mathrm{B}}$, entering the Planck function $\Pi(\omega, T) =
\hbar\omega/ [\mathrm{exp}(\hbar\omega/(k_{\mathrm{B}} T)) - 1]$ (and
it has been assumed, without loss of generality, that $T_{\mathrm{B}}
> T_{\mathrm{A}}$ so $P > 0$), and the spectral function
$\Phi(\omega)$, which can be enhanced by changing material and
geometric properties through the creation of resonances and changes in
the electromagnetic density of states. In particular, nanostructuring
metallic surfaces or polar dielectrics makes it possible to shift
resonant frequencies from the visible or far infrared into the
infrared, such that the peak of the spectrum $\Phi$ may coincide with
the peak of the Planck distribution near room temperature. It remains
an open question, however, to what extent the peak value of $\Phi$
itself may be enhanced through appropriate geometric and material
choices, as well as what such optimal structures should be.

Previous attempts at deriving bounds on RHT have primarily focused on
extended media~\cite{PendryJPCM1999, BimontePRA2009, BiehsPRL2010,
  BenAbdallahPRB2010}, showing that at least for translationally
invariant structures, $\Phi$ can be expressed as the trace of a
``transmission'' matrix whose singular values (corresponding to
evanescent Fourier modes) each contribute a finite flux, bounded above
by a Landauer limit in analogy with conduction~\cite{Datta1995,
  KlocknerPRB2016}. Aside from being restricted to planar geometries,
these bounds turn out to be either
pessimistic~\cite{BenAbdallahPRB2010}, ignoring the large densities of
states that can arise in nanostructured and low-loss materials, or too
optimistic~\cite{PendryJPCM1999, BiehsPRL2010}, ignoring any
constraints imposed by Maxwell's equations and assuming instead that
all such Fourier modes, up to an unrealistic cutoff on the order of
the atomic scale, can saturate the flux~\cite{PendryJPCM1999}. From a
design perspective, Landauer limits present a hurdle as they rely on
ad-hoc estimates of the number and relative contribution of radiative
modes/channels, which depend on specific material and geometric
features. More recent works have derived complementary material limits
on electromagnetic absorption in subwavelength
regimes~\cite{MillerOE2016}, showing that absorbed power in a medium
of susceptibility $\chi$ increases in proportion to an ``inverse
resistivity'' figure of merit, $|\chi|^2/\Im\chi$, in principle
diverging with increasing indices of refraction and decreasing
dissipation. Saturation of these bounds for a subwavelength absorber
in the quasistatic regime can generally be achieved through the strong
polarization currents arising in resonant media supporting surface
plasmon or phonon polaritons. These arguments have been extended to
near-field RHT~\cite{MillerPRL2015} by exploiting energy conservation
and reciprocity, finding the upper bound of $\Phi$ at a polariton
resonance to scale quadratically with $|\chi|^2/\Im\chi$,
corresponding to enhanced absorption and emission in both
objects. While near-field RHT between dipolar objects can attain these
bounds in a dilute limit, such a universal scaling has yet to be
observed in large-area structures. This na\"{i}vely suggests room for
improvement in $\Phi$ through nanostructuring via enhancements in the
density of states or equivalently, via saturation of modal
contributions, yet trial-and-error explorations and optimization
procedures~\cite{JinOE17, FernandezHurtadoPRL2017} have failed to
produce nanostructured geometries that bridge this gap, leading to the
alternative possibility that existing bounds are too loose.

In this paper, we derive new algebraic bounds on RHT, valid in the
near-, mid-, and far-field regimes, through analysis of the singular
value decompositions of relevant response quantities. In contrast to
prior limits, our bounds incorporate constraints imposed by material
losses and multiple scattering, and are therefore tighter; moreover,
they are formulated to be independent of object shapes while
simultaneously accounting for finite size effects. In particular,
every channel of energy transmission is shown to be generally
prohibited from saturating its Landauer limit, in contrast to
predictions based on modal decompositions~\cite{PendryJPCM1999,
  BimontePRA2009, BiehsPRL2010, BenAbdallahPRB2010} that neglect
material properties and are most applicable in the ray optics
regime. Furthermore, the growth of RHT with decreasing material
dissipation is shown to be strongly limited by radiative losses, in
contrast to predictions based on energy-conservation limits to
material response~\cite{MillerPRL2015} that neglect finite-size
scattering effects and are thus most applicable in the quasistatic
regime. Finally, we discuss the difficulties in optimizing each
individual energy transmission channel and show how more restricted
bounds may be obtained by approximating each body to maximally absorb
all incident fields in isolation. In upcoming papers closely related
to this one, we apply these bounds to various scenarios of interest,
providing predictions of the maximum RHT achievable in compact and
extended geometries~\cite{VenkataramARXIV2019}, and deriving related
bounds on far-field thermal emission from single bodies in
isolation~\cite{MoleskyARXIV2019B}.

\section{Heat transfer definitions}

\begin{figure}[t!]
\centering
\includegraphics[width=0.9\columnwidth]{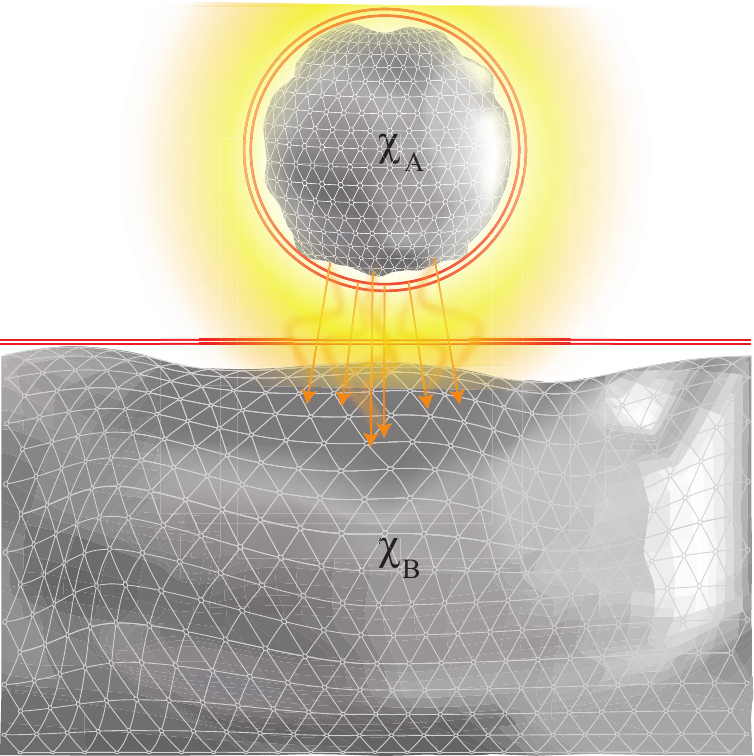}
\caption{Two bodies labeled A and B exchange heat in vacuum. Each body
  could be compact or of infinite extent in at least one spatial
  dimension, and for given susceptibilities $\chi_{p}$, the optimal
  structures may be quite complicated, but the upper bounds, which
  depend on $\zeta_{p} = |\chi_{p}|^{2} / \Im(\chi_{p})$, can be
  evaluated in simpler bounding domains that enclose each object while
  respecting any other design constraints present.}
\label{fig:schematic}
\end{figure}

For two bodies A and B in vacuum [\figref{schematic}], the spectral
function $\Phi$ appearing in~\eqref{integratedRHT} represents the
average power absorbed in B due to fluctuating current sources in A %
, depicted in
% \figref{scat-schem},
and is reciprocal (invariant under interchange of A and B). Using
formal operator notation, this average absorbed power can be written
in terms of the susceptibilities $\VV_{p}$, the vacuum Maxwell Green's
function $\Gvac_{pq}$, and scattering T-operators $\TT_{p}$, for
$p, q \in \{\mathrm{A}, \mathrm{B}\}$. Each susceptibility is written
as $\VV_{p} = \chi_{p} \II_{p}$, where each $\chi_{p}$ is assumed to
be homogeneous, local, and isotropic. The vacuum Maxwell Green's
function $\Gvac$ solves
$[(c/\omega)^{2} \nabla \times (\nabla \times) - \II]\Gvac = \II$ in
all space, and its blocks are denoted as $\Gvac_{pq}$ for sources in
body $q$ propagating fields to body $p$. Finally, the T-operators
$\TT_{p} = (\VV_{p}^{-1} - \Gvac_{pp})^{-1}$ represent the total
induced polarization moment in body $p$ due to a localized field of
unit magnitude incident upon it. All of these quantities are
reciprocal, so they are equal to their unconjugated transposes in
position space: $\VV_{p} = \VV_{p}^{\top}$,
$\TT_{p} = \TT_{p}^{\top}$, and $\Gvac_{pq} =
(\Gvac_{qp})^{\top}$. This means that Hermitian conjugation is
equivalent to complex conjugation:
$\VV_{p}^{\dagger} = \VV_{p}^{\star}$,
$\TT_{p}^{\dagger} = \TT_{p}^{\star}$, and
$(\Gvac_{pq})^{\dagger} = \GG_{qp}^{\mathrm{vac}\star}$. Additionally,
all of these quantities depend on frequency $\omega$, though this will
be suppressed for the sake of notational brevity.

Given these definitions and relations (see appendices for more
details), the RHT spectrum can be written as~\cite{KrugerPRB2012}
\begin{multline} \label{eq:exactPhibothT}
  \Phi = \frac{2}{\pi}
  \tr[\TT_{\mathrm{B}}^{\star} (\II_{\mathrm{B}} -
  \GG_{\mathrm{BA}}^{\mathrm{vac}\star} \TT_{\mathrm{A}}^{\star}
  \GG_{\mathrm{AB}}^{\mathrm{vac}\star} \TT_{\mathrm{B}}^{\star})^{-1}
  \Im(\VV_{\mathrm{B}}^{-1\star}) \times \\ \TT_{\mathrm{B}} (\II_{\mathrm{B}} -
  \Gvac_{\mathrm{BA}} \TT_{\mathrm{A}} \Gvac_{\mathrm{AB}}
  \TT_{\mathrm{B}})^{-1} \times \\ \Gvac_{\mathrm{BA}} \TT_{\mathrm{A}}
  \Im(\VV_{\mathrm{A}}^{-1\star}) \TT_{\mathrm{A}}^{\star}
  \GG_{\mathrm{AB}}^{\mathrm{vac}\star}],
\end{multline}
where $\Im(\mathbb{A}) = (\mathbb{A} - \mathbb{A}^{\star})/(2\im)$ and
$\asym(\mathbb{A}) = (\mathbb{A} - \mathbb{A}^{\dagger})/(2\im)$ for
any operator $\mathbb{A}$; if $\mathbb{A}$ is reciprocal, then
$\asym(\mathbb{A}) = \Im(\mathbb{A})$. This expression is manifestly
reciprocal in A and B, and treats the T-operators of A and B on an
equal footing, linked only by the Green's function
$\Gvac_{\mathrm{BA}}$ propagating fields in vacuum from one body to
the other. However, it is possible to write this spectrum more
suggestively in terms of operator combinations that hide this
reciprocity in order to more strongly link this expression to absorbed
and emitted powers. In particular, $\Phi$ may be rewritten as
\begin{equation} \label{eq:exactPhiTAYB}
  \Phi = \frac{2}{\pi}
  \tr[\YY_{\mathrm{B}}^{\star} \Im(\VV_{\mathrm{B}}^{-1\star})
  \YY_{\mathrm{B}} \Gvac_{\mathrm{BA}} \TT_{\mathrm{A}}
  \Im(\VV_{\mathrm{A}}^{-1\star}) \TT_{\mathrm{A}}^{\star}
  \GG_{\mathrm{AB}}^{\mathrm{vac}\star}],
\end{equation}
in terms of the reciprocal operator
$\YY_{\mathrm{B}} = \TT_{\mathrm{B}} \Ss^{\mathrm{A}}_{\mathrm{B}}$,
which is in turn written in terms of the scattering operator
$\Ss^{\mathrm{A}}_{\mathrm{B}} = (\II_{\mathrm{B}} -
\Gvac_{\mathrm{BA}} \TT_{\mathrm{A}} \Gvac_{\mathrm{AB}}
\TT_{\mathrm{B}})^{-1}$: essentially, $\YY_{\mathrm{B}}$ is a new
``dressed T-operator'' describing absorption and scattering in B in
the presence of A, just as the bare T-operators $\TT_{p}$ describe
absorption and scattering from each body
$p \in \{\mathrm{A}, \mathrm{B}\}$ in isolation.

Having assumed the susceptibility in each body $p \in \{\mathrm{A},
\mathrm{B}\}$ to be homogeneous, uniform, and isotropic, makes it
possible to identify $\Im(\VV_{p}^{-1\star}) =
\frac{\Im(\chi_{p})}{|\chi_{p}|^{2}} \II_{p}$. For convenience, we
denote $\zeta_{p} = \frac{|\chi_{p}|^{2}}{\Im(\chi_{p})}$ as the
material response factor. Using this, we write the RHT spectrum as
\begin{equation} \label{eq:exactPhiTAYBFrob}
  \Phi = \frac{2}{\pi\zeta_{\mathrm{A}} \zeta_{\mathrm{B}}} \left\lVert \YY_{\mathrm{B}} \Gvac_{\mathrm{BA}} \TT_{\mathrm{A}} \right\rVert_{\mathrm{F}}^{2}
\end{equation}
where
$\left\lVert \mathbb{A} \right\rVert_{\mathrm{F}}^{2} =
\tr(\mathbb{A}^{\dagger} \mathbb{A})$ denotes the Frobenius norm for
any operator $\mathbb{A}$.

As we show in the appendices, the RHT spectrum may alternatively be
written as
\begin{equation}
  \Phi = \frac{2}{\pi} \left\lVert \mathbb{Q} \right\rVert_{\mathrm{F}}^{2}
\end{equation}
in terms of the transmission operator $\mathbb{Q} =
\Im(\VV_{\mathrm{B}})^{1/2} \GG_{\mathrm{BA}}
\Im(\VV_{\mathrm{A}})^{1/2}$, which in turn depends on the total
Green's function $\GG_{\mathrm{BA}} = \VV_{\mathrm{B}}^{-1}
\TT_{\mathrm{B}} \Ss^{\mathrm{A}}_{\mathrm{B}} \Gvac_{\mathrm{BA}}
\TT_{\mathrm{A}} \VV_{\mathrm{A}}^{-1}$ connecting dipole sources in
body A to total fields in body B and accounting for multiple
scattering to all orders within and between both bodies. This obeys a
Landauer limit, as the singular values of $\mathbb{Q}^{\dagger}
\mathbb{Q}$ do not exceed $1/4$, so including the prefactor $2/\pi$,
the contribution of each mode/channel in the trace expression to
$\Phi$ does not exceed $\frac{1}{2\pi}$. We will shortly explain the
conditions under which the Landauer bounds for each of these
contributions may be saturated.

\section{Singular value bounds}

Each of the operators $\YY_{\mathrm{B}}$, $\Gvac_{\mathrm{BA}}$, and
$\TT_{\mathrm{A}}$ may be decomposed into their respective singular
values and vectors. In general, the singular vectors of each of these
operators will not bear any relation to one another. However, we prove
in the appendices that an upper bound to $\Phi$ can be found by
ensuring that the right singular vectors of $\YY_{\mathrm{B}}$ and the
left singular vectors of $\TT_{\mathrm{A}}$ are respectively the same
as the left and right singular vectors of
$\Gvac_{\mathrm{BA}}$. Intuitively, these conditions can be
interpreted as representing perfect coupling and propagation of fields
due to sources in A which are perfectly absorbed in body B (while
accounting for multiple scattering from the presence of body
A). Formally, we write the statements
\begin{align}
  \YY_{\mathrm{B}} &= \sum_{i} y_{i} \ket{\vec{b}_{i}^{\star}} \bra{\vec{b}_{i}} \nonumber \\
  \Gvac_{\mathrm{BA}} &= \sum_{i} g_{i} \ket{\vec{b}_{i}} \bra{\vec{a}_{i}} \nonumber \\
  \TT_{\mathrm{A}} &= \sum_{i} \tau_{i} \ket{\vec{a}_{i}}
                     \bra{\vec{a}_{i}^{\star}} \label{eq:opSVD}
\end{align}
where as a reminder, the singular values $\tau_{i}$, $g_{i}$, and
$y_{i}$ are nonnegative real numbers, and the vectors
$\ket{\vec{a}_{i}}$ and $\ket{\vec{b}_{i}}$ are orthonormal among
themselves so $\bracket{\vec{a}_{i}, \vec{a}_{j}} = \delta_{ij}$ and
$\bracket{\vec{b}_{i}, \vec{b}_{j}} = \delta_{ij}$. Our explicit
conjugation of $\ket{\vec{b}_{i}^{\star}}$ in the definition of
$\YY_{\mathrm{B}}$ and of $\bra{\vec{a}_{i}^{\star}}$ in the
definition of $\TT_{\mathrm{A}}$ ensure that they still respect
reciprocity. This allows us to write a bound on the RHT spectrum as
\begin{equation} \label{eq:Phiboundallsingvals}
  \Phi = \frac{2}{\pi\zeta_{\mathrm{A}} \zeta_{\mathrm{B}}} \sum_{i}
  (y_{i} g_{i} \tau_{i})^{2}
\end{equation}
which is maximized when $y_{i}$ and $\tau_{i}$ attain their maximum
values as functions of $g_{i}$; the singular values $g_{i}$ depend
only on the geometry and topology of the system as they enter
$\Gvac_{\mathrm{BA}}$, not the material properties, so we take those
as fixed.

As we show in the appendices, nonnegative far-field scattering from
the combined system of bodies A and B together implies that the
operator $\Im(\YY_{\mathrm{B}}) - \YY_{\mathrm{B}}
\left(\frac{1}{\zeta_{\mathrm{B}}} \II_{\mathrm{B}} +
\frac{1}{\zeta_{\mathrm{A}}} \Gvac_{\mathrm{BA}} \TT_{\mathrm{A}}
\TT_{\mathrm{A}}^{\star} \GG_{\mathrm{AB}}^{\mathrm{vac}\star}\right)
\YY_{\mathrm{B}}^{\star}$ is not only Hermitian but also
positive-semidefinite. This implies that if $\YY_{\mathrm{B}}$ had a
nontrivial Hermitian part, any bounds on the singular values $y_{i}$
arising from this total nonnegative far-field scattering constraint
must be more restrictive than if $\YY_{\mathrm{B}}$ were purely
anti-Hermitian, meaning $\YY_{\mathrm{B}} =
\im\Im(\YY_{\mathrm{B}})$. Thus, we take $\YY_{\mathrm{B}}$ to be
anti-Hermitian, which also means that $\bracket{\vec{b}_{j},
  \vec{b}_{j}^{\star}} = \im$ for every channel $j$. Additionally, we
note that $\Gvac_{\mathrm{BA}} \TT_{\mathrm{A}}
\TT_{\mathrm{A}}^{\star} \GG_{\mathrm{AB}}^{\mathrm{vac}\star} =
\sum_{i} g_{i}^{2} \tau_{i}^{2} \ket{\vec{b}^{(i)}}
\bra{\vec{b}^{(i)}}$ follows from our singular value decompositions,
so it remains to be proved that the operator $\sum_{j}
\frac{y_{j}}{2\im} \left(\ket{\vec{b}_{j}^{\star}}\bra{\vec{b}_{j}} -
\ket{\vec{b}_{j}}\bra{\vec{b}_{j}^{\star}}\right) - \sum_{j} y_{j}^{2}
\left(\frac{1}{\zeta_{\mathrm{B}}} + \frac{1}{\zeta_{\mathrm{A}}}
(g_{j} \tau_{j})^{2}\right) \ket{\vec{b}_{j}^{\star}}
\bra{\vec{b}_{j}^{\star}}$ is positive-semidefinite. To do this, we
take the inner product of this operator with
$\ket{\vec{b}_{i}^{\star}}$ on the right and
$\bra{\vec{b}_{i}^{\star}}$ on the left for a single fixed index $i$
(not to be confused with the imaginary unit), which yields the
condition $y_{i} - y_{i}^{2} \left(\frac{1}{\zeta_{\mathrm{B}}} +
\frac{1}{\zeta_{\mathrm{A}}} (g_{i} \tau_{i})^{2}\right) \geq 0$. This
inequality is saturated to give the largest possible $y_{i} =
\frac{\zeta_{\mathrm{B}}}{1 +
  \frac{\zeta_{\mathrm{B}}}{\zeta_{\mathrm{A}}} (g_{i} \tau_{i})^{2}}$
in terms of $g_{i} \tau_{i}$, which in turn gives the largest possible
flux,
\begin{equation*}
  \Phi = \frac{2\zeta_{\mathrm{B}}}{\pi\zeta_{\mathrm{A}}} \sum_{i}
  \frac{(g_{i} \tau_{i})^{2}}{\left[1 +
      \frac{\zeta_{\mathrm{B}}}{\zeta_{\mathrm{A}}} (g_{i}
      \tau_{i})^{2}\right]^{2}}
\end{equation*}
in terms of $g_{i} \tau_{i}$.

The above expression is maximized for each channel $i$, and for the
sum in turn, if $\frac{\zeta_{\mathrm{B}}}{\zeta_{\mathrm{A}}} (g_{i}
\tau_{i})^{2} = 1$. This is solved to yield $\tau_{i} =
\frac{1}{\sqrt{\frac{\zeta_{\mathrm{B}}}{\zeta_{\mathrm{A}}}} g_{i}}$,
and plugging this into the expression for $y_{i}$ yields $y_{i} =
\frac{\zeta_{\mathrm{B}}}{2}$, while plugging this into the expression
for $\Phi$ yields a contribution $\frac{2}{\pi\zeta_{\mathrm{A}}
  \zeta_{\mathrm{B}}} (y_{i} g_{i} \tau_{i})^{2} = \frac{2}{\pi}
\times \frac{1}{4} = \frac{1}{2\pi}$. We interpret this to mean that
to obtain optimal heat transfer, the T-operator of body A in isolation
must be engineered in a way that depends on the presence of body B,
due to both the presence of the material enhancement factor
$\zeta_{\mathrm{B}}$ and the dependence on the singular values $g_{i}$
of $\Gvac_{\mathrm{BA}}$ propagating electromagnetic fields in vacuum
from A to B, though no higher-order scattering effects come into
play. In turn, the expression $y_{i} = \frac{\zeta_{\mathrm{B}}}{2}$
means that the effective T-operator of body B dressed by scattering
from body A must actually exhibit maximal \emph{scattering}, and not
absorption, in the presence of body A~\cite{MillerOE2016}, though it
is more difficult to extract information about the implications for
the T-operator of body B in isolation; we clarify that maximal
scattering includes both far-field scattering from body B in the
presence of body A, as well as absorption from body A in the presence
of body B. If these two conditions can be met simultaneously for the
given channel $i$, which is effectively a rate-matching condition
relating the absorption and scattering rates of each body in the
presence of the other, then the contribution $\frac{1}{2\pi}$ is
exactly the Landauer upper bound on energy transmission.

However, this analysis has so far ignored constraints on
$\tau_{i}$. In particular, nonnegativity of far-field scattering from
body A in isolation requires that $\Im(\TT_{\mathrm{A}}) -
\frac{1}{\zeta_{\mathrm{A}}} \TT_{\mathrm{A}}
\TT_{\mathrm{A}}^{\star}$ be a positive-semidefinite Hermitian
operator~\cite{MillerOE2016}. Following similar steps as above, we may
see that for the purposes of bounding $\tau_{i}$, the loosest
constraints on $\tau_{i}$ arise if $\TT_{\mathrm{A}}$ is purely
anti-Hermitian, so taking the inner product of the above operator with
respect to $\ket{\vec{a}_{i}}$ on the right and $\bra{\vec{a}_{i}}$ on
the left yields the condition $\tau_{i} \leq \zeta_{\mathrm{A}}$. The
expression $\tau_{i} =
\frac{1}{\sqrt{\frac{\zeta_{\mathrm{B}}}{\zeta_{\mathrm{A}}}} g_{i}}$
is consistent with this inequality only if the inequality $g_{i} \geq
\frac{1}{\sqrt{\zeta_{\mathrm{A}} \zeta_{\mathrm{B}}}}$ also holds for
the given index $i$. If not, and instead $g_{i} <
\frac{1}{\sqrt{\zeta_{\mathrm{A}} \zeta_{\mathrm{B}}}}$, then
$\tau_{i} = \zeta_{\mathrm{A}}$ must be used to maximize the
contribution at index $i$ to $\Phi$, saturating the bound on the
response of A in isolation. This yields $y_{i} =
\frac{\zeta_{\mathrm{B}}}{1 + \zeta_{\mathrm{A}} \zeta_{\mathrm{B}}
  g_{i}^{2}} \geq \frac{\zeta_{\mathrm{B}}}{2}$ and a contribution of
$\frac{2}{\pi} \frac{\zeta_{\mathrm{A}} \zeta_{\mathrm{B}}
  g_{i}^{2}}{(1 + \zeta_{\mathrm{A}} \zeta_{\mathrm{B}}
  g_{i}^{2})^{2}} \leq \frac{1}{2\pi}$ to $\Phi$. We interpret this to
mean that if the singular value $g_{i}$ of $\Gvac_{\mathrm{BA}}$ falls
below a threshold involving the two material enhancement factors, then
the optimal T-operator of body A in isolation corresponds to maximal
absorption, the optimal effective T-operator of body B dressed by body
A evinces the effects of multiple scattering with A, and the
contribution to $\Phi$ similarly shows the effects of multiple
scattering between the two bodies and is unable to saturate the
Landauer bound for that channel.

To summarize, the bound on RHT may be written as
\begin{multline} \label{eq:Phiboundrhosingvals}
  \Phi_{\mathrm{opt}} =
  \sum_{i} \Big[\frac{1}{2\pi} \Theta(\zeta_{\mathrm{A}}
    \zeta_{\mathrm{B}} g_{i}^{2} - 1) \\ + \frac{2}{\pi}
    \frac{\zeta_{\mathrm{A}} \zeta_{\mathrm{B}} g_{i}^{2}}{(1 +
      \zeta_{\mathrm{A}} \zeta_{\mathrm{B}} g_{i}^{2})^{2}}
    \Theta(1 - \zeta_{\mathrm{A}} \zeta_{\mathrm{B}}
    g_{i}^{2})\Big]
\end{multline}
where $\Theta$ is the Heaviside step function. As the material
response (encoded in $\zeta_{\mathrm{A}} \zeta_{\mathrm{B}}$)
increases, progressively more channels may saturate the Landauer limit
per channel, so that the Landauer limit (summed over all channels) is
reached asymptotically as
$\zeta_{\mathrm{A}} \zeta_{\mathrm{B}} \to \infty$. However, the rate
at which this divergence occurs depends on the general topology of the
problem, as that determines how the singular values $g_{i}$ of
$\Gvac_{\mathrm{BA}}$ depend on the index $i$.

We emphasize that while the singular values $g_{i}$ of
$\Gvac_{\mathrm{BA}}$ are technically restricted to the domains of the
objects to give the tightest bound on heat transfer, such a
restriction is less than ideal given the explicit dependence on the
shapes of the objects. However, as we prove in the appendices, the
singular values $g_{i}$ of $\Gvac_{\mathrm{BA}}$ are domain monotonic,
meaning that they increase monotonically as the volumes of regions A
and B increase; consequently, $\Phi_{\mathrm{opt}}$ is domain
monotonic, as it is monotonically non-decreasing with respect to
$g_{i}$ for each $i$. Separately from this, the regions containing
only the material degrees of freedom of each body can be replaced by
larger regions that fully enclose each body, as the T-operators of
each body will commute with projections into the smaller subspaces
corresponding to the actual material degrees of freedom. Thus, these
bounds can be slightly loosened to be independent of body shapes, and
can then be evaluated subject to constraints on topology and domain
volumes as determined by the desired application [\figref{schematic}],
e.g. ellipsoids with prescribed aspect ratios or films of prescribed
thicknesses representative of compact or extended geometries,
respectively.  Essentially, the effective rank of
$\Gvac_{\mathrm{BA}}$, which determines the number of modes that could
participate in RHT, is largely determined by the size and topology of
the choice of bounding surface, which represents a general and
fundamental geometric constraint on the bounds of RHT analogous to the
general material constraints imposed by $\zeta_{p}$ for each body $p
\in \{\mathrm{A}, \mathrm{B}\}$.

\section{Comparison to alternative bounds}

The bound for the RHT spectrum $\Phi$ in~\eqref{Phiboundrhosingvals}
may be compared to a number of other bounds. Strictly speaking, this
bound is not necessarily the tightest general bound that could be
formulated. In particular, using the relation $\TT_{\mathrm{A}}^{-1} =
\VV_{\mathrm{A}}^{-1} - \Gvac_{\mathrm{AA}}$ allows for
writing~\eqref{exactPhiTAYB} in terms of $\TT_{\mathrm{A}}$ and
$\Im(\Gvac_{\mathrm{AA}})$ without reference to
$\VV_{\mathrm{A}}$. Such a procedure, in analogy with bounds on
thermal emission which we detail in an upcoming
paper~\cite{MoleskyARXIV2019B}, would more explicitly capture
far-field radiative losses from bodies of finite size, which becomes
more relevant at large separations where such losses may compete with
RHT itself. However, as we show in the appendices, we find the
resulting bound to be intractable, requiring self-consistent solution
of systems of nonlinear equations to find the optimal singular values
of $\TT_{\mathrm{A}}$. Therefore, we do not further consider such a
bound, and henceforth refer only to~\eqref{Phiboundrhosingvals}.

With respect to prior work, the most obvious point of comparison is
the Landauer bound~\cite{PendryJPCM1999}, namely
\begin{equation}
  \Phi_{\mathrm{L}} = \sum_{i} \frac{1}{2\pi},
\end{equation}
which simply depends on the number of modes participating in RHT,
without any reference to separation, geometric constraints, multiple
scattering, or even material constraints, let alone their interplay;
consequently, in contrast to our bounds, there is no metric to
evaluate how many participating modes can actually saturate the limit
$\frac{1}{2\pi}$. Even modal analyses that technically do not
necessarily assume saturation of the Landauer limits for every
mode~\cite{PendryJPCM1999, BimontePRA2009, BiehsPRL2010,
  BenAbdallahPRB2010} tend to neglect material effects, so the purely
geometric arguments are valid only in the ray-optical regime where
blackbody limits are reproduced.

An alternative bound can be derived by modifying the form
of~\eqref{Phiboundrhosingvals} such that the contributions from
saturating the singular value bounds for $\tau_{i}$ are used for all
indices $i$, not only those where
$\zeta_{\mathrm{A}} \zeta_{\mathrm{B}} g_{i}^{2} < 1$. We may write
this as
\begin{equation}
  \Phi_{\mathrm{sc}} = \sum_{i} \frac{2}{\pi}
  \frac{\zeta_{\mathrm{A}} \zeta_{\mathrm{B}} g_{i}^{2}}{(1 +
    \zeta_{\mathrm{A}} \zeta_{\mathrm{B}} g_{i}^{2})^{2}}
\end{equation}
and term this the ``scalar approximation'' with the subscript ``sc''
because (as shown in the appendices) it can equivalently be derived by
plugging into~\eqref{exactPhibothT} the T-operators $\TT_{p} =
\im\tau_{p} \II_{p}$ for each body $p \in \{\mathrm{A}, \mathrm{B}\}$,
corresponding to uniform singular values $\tau_{p}$ for each
$\TT_{p}$; in the appendices, we further derive that this bound on RHT
is maximal when each body exactly satisfies the condition for perfect
isolated absorption~\cite{MillerOE2016} for every incident field,
meaning $\tau_{p} = \zeta_{p}$. As each contribution to
$\Phi_{\mathrm{sc}}$ depends on the singular values $g_{i}$ in a
nonmonotonic way, domain monotonicity is not obvious, though arguments
based on the embedding of the T-operators in larger spaces allow for
consideration of general bounding surfaces as geometric constraints,
making this a useful bound restricted to the class of T-operators with
uniform singular values. However, as we show in an upcoming paper,
$\Phi_{\mathrm{sc}}$ overall can be shown to be domain monotonic in
the near-field.

Another such bound neglects multiple scattering effects between the
two bodies, and can be derived in several ways. As we show in the
appendices, this bound follows from the scalar approximation
$\Phi_{\mathrm{sc}}$ by neglecting the scattering operator
$\Ss^{\mathrm{A}}_{\mathrm{B}} \to \II_{\mathrm{B}}$, or alternatively
by ignoring the denominators
$(1 + \zeta_{\mathrm{A}} \zeta_{\mathrm{B}} g_{i}^{2})^{2}$ in
$\Phi_{\mathrm{sc}}$ encoding the effects of multiple scattering
within each channel. Regardless of the particular derivation, the end
result may be written as
\begin{equation}
  \Phi_{\mathrm{Born}} = \sum_{i} \frac{2}{\pi} \zeta_{\mathrm{A}} \zeta_{\mathrm{B}} g_{i}^{2}
\end{equation}
which we term the ``Born bound'' as its neglect of multiple scattering
effects between the two bodies is characteristic of a Born
approximation to the full solution of Maxwell's equations. The domain
monotonicity of this bound trivially follows from that of $g_{i}$ for
each channel $i$, and the validity of the embedding argument with
respect to $\TT_{p}$ still holds, meaning that $\Phi_\mathrm{Born}$ is
also a useful bound for the RHT spectrum when considering bounding
surfaces of arbitrary size and topology. We note that Miller et
al~\cite{MillerPRL2015} equivalently derive this bound by limiting the
singular values of $\YY_{\mathrm{B}}$ such that only far-field
scattering of body B in the presence of body A needs to be
nonnegative, meaning $\Im(\YY_{\mathrm{B}}) - \YY_{\mathrm{B}}^{\star}
\Im(\VV_{\mathrm{B}}^{-1\star}) \YY_{\mathrm{B}}$ should be
positive-semidefinite; this leads to the bound $y_{i} \leq
\zeta_{\mathrm{B}}$, so maximization of~\eqref{Phiboundallsingvals}
subject to that constraint as well as $\tau_{i} \leq
\zeta_{\mathrm{A}}$ simply requires saturation of both of these
constraints. However, we have shown that positive-semidefiniteness of
$\Im(\YY_{\mathrm{B}}) - \YY_{\mathrm{B}}^{\star}
\Im(\VV_{\mathrm{B}}^{-1\star}) \YY_{\mathrm{B}}$, corresponding to
nonnegative scattering from body B in the presence of body A, is a
looser constraint on $y_{i}$ than nonnegativity of scattering from the
system as a whole, corresponding to positive-semidefiniteness of
$\Im(\YY_{\mathrm{B}}) - \YY_{\mathrm{B}}
\left(\Im(\VV_{\mathrm{B}}^{-1\star}) + \Gvac_{\mathrm{BA}}
\TT_{\mathrm{A}} \Im(\VV_{\mathrm{A}}^{-1\star})
\TT_{\mathrm{A}}^{\star} \GG_{\mathrm{AB}}^{\mathrm{vac}\star}\right)
\YY_{\mathrm{B}}^{\star}$, and that only the latter produces a bound
on $y_{i}$ that explicitly accounts for multiple scattering and
absorption losses through body A. The Born bound not only neglects
multiple scattering but also ignores far-field radiative losses, which
means that while it is technically a looser bound, its usefulness is
limited to deeply subwavelength systems exchanging heat in the
near-field.

We are now in a position to relate $\Phi_{\mathrm{opt}}$,
$\Phi_{\mathrm{sc}}$, $\Phi_{\mathrm{Born}}$, and $\Phi_{\mathrm{L}}$
to each other via inequalities. First, the discussion
of~\eqref{Phiboundrhosingvals} immediately makes clear that because
some of the contributions to $\Phi_{\mathrm{opt}}$ equal corresponding
contributions to $\Phi_{\mathrm{L}}$ while all of the others equal
corresponding contributions to $\Phi_{\mathrm{sc}}$, then it must be
the case that $\Phi_{\mathrm{sc}} \leq \Phi_{\mathrm{opt}} \leq
\Phi_{\mathrm{L}}$. Additionally, because the contributions to
$\Phi_{\mathrm{opt}}$ from each channel $i$ never exceed
$\frac{1}{2\pi}$, and in particular because the saturation of the
Landauer bound for each channel occurs for $\zeta_{\mathrm{A}}
\zeta_{\mathrm{B}} g_{i}^{2} \geq 1$, then it must be the case that
$\Phi_{\mathrm{sc}} \leq \Phi_{\mathrm{opt}} \leq
\Phi_{\mathrm{Born}}$, because in the expression for
$\Phi_{\mathrm{Born}}$, the contributions $\zeta_{\mathrm{A}}
\zeta_{\mathrm{B}} g_{i}^{2}$ will always trivially exceed $1/4$ when
$\zeta_{\mathrm{A}} \zeta_{\mathrm{B}} g_{i}^{2} \geq 1$, and will
always exceed $\frac{\zeta_{\mathrm{A}} \zeta_{\mathrm{B}}
  g_{i}^{2}}{(1 + \zeta_{\mathrm{A}} \zeta_{\mathrm{B}}
  g_{i}^{2})^{2}}$ when $\zeta_{\mathrm{A}} \zeta_{\mathrm{B}}
g_{i}^{2} < 1$. In general, it is not possible to write an inequality
relation between $\Phi_{\mathrm{Born}}$ and $\Phi_{\mathrm{L}}$ in all
situations, because $\Phi_{\mathrm{Born}}$ may have some contributions
$\frac{2}{\pi} \zeta_{\mathrm{A}} \zeta_{\mathrm{B}} g_{i}^{2}$ which
fall above or below $\frac{1}{2\pi}$, and the geometry determining
$g_{i}$ would have to be known in order to know how many fall above or
below. Thus, we may write the overall inequalities as
\begin{equation}
  \Phi_{\mathrm{sc}} \leq \Phi_{\mathrm{opt}} \leq
  \Phi_{\mathrm{Born}}, \Phi_{\mathrm{L}}.
\end{equation}

\section{Concluding remarks}

We have determined bounds for the RHT spectrum $\Phi$ based purely on
algebraic arguments. In particular, we show that there is a tension
between optimizing transmission channels and material/geometric
constraints placed on each object in isolation as well as in in the
presence of the other; as a result, some but not all channels can
saturate previously derived Landauer bounds, while others are
restricted by the aforementioned constraints. By virtue of domain
monotonicity, these bounds can be applied in a shape-independent
manner, so while they can be evaluated analytically in highly
symmetric bounding surfaces, they can just as easily be evaluated
numerically in more complicated domains depending on specific design
constraints [\figref{schematic}]. Similarly, the dependence on the
material response factor $\zeta = |\chi|^{2} / \Im\chi$ does not make
explicit reference to a particular frequency or material model. In
comparison, the Landauer bounds yield overly optimistic predictions,
while choosing a scalar response for each object corresponding to
maximal absorption of every incident field in isolation yields overly
pessimistic predictions. Additionally, we find that previous work by
Miller et al~\cite{MillerPRL2015} also yields overly optimistic
predictions compared to our current bounds, because those derivations
neglect the geometric effects of multiple scattering between the two
bodies and consequently overestimate the optimal response of one body
in the presence of the other. We point out that while our bounds are
always tighter than Landauer and Born limits for any given bounding
domain, they say nothing about which domains may yield the tightest
per-volume limits given material constraints, or whether they may in
fact be attained by physically realizable structures.In summary, while
Born and Landauer limits are technically upper bounds on RHT, their
neglect of multiple scattering and material losses, respectively,
render them loose compared to the bounds presented here.

We emphasize that in contrast to Born limits~\cite{MillerPRL2015}, our
bounds are valid from the near-field all the way through the far-field
for bodies of arbitrary size, as we never needed to exploit
nonretarded or quasistatic approximations. We also point out that in the
near-field, $\Gvac_{\mathrm{BA}}$ is real-valued in position space, so
arguments involving the anti-Hermiticity of $\Gvac_{\mathrm{BA}}
\TT_{\mathrm{A}} \Gvac_{\mathrm{AB}}$ imply the anti-Hermiticity of
$\TT_{\mathrm{A}}$, as $\Im(\Gvac_{\mathrm{BA}} \TT_{\mathrm{A}}
\Gvac_{\mathrm{AB}}) \to \Gvac_{\mathrm{BA}} \Im(\TT_{\mathrm{A}})
\Gvac_{\mathrm{AB}}$. In this context, Born bounds only yield sensible
results in the near-field, as they tend to diverge beyond the
near-field regime for objects of infinite extent in at least one
spatial dimension faster than expected.

In a complementary paper~\cite{VenkataramARXIV2019}, we analyze these
bounds in the near-field in specific geometries of interest,
particularly dipolar and extended (infinite area) bodies. We take
particular note of the scalar approximation because, though
representing a restricted version of our general bounds and resulting
in comparably pessimistic predictions, it may prove challenging if not
impossible to optimize every individual channel to saturate its
Landauer limit when the material constraints are not saturated. While
this mere observation does not preclude the possibility of larger RHT
(as allowed by our general bounds), from a design perspective it might
prove more feasible to consider bodies capable of maximally absorbing
every incident field in isolation, as could occur near a polaritonic
resonance. Moreover, as we show in this follow-up work, the simplicity
of the scalar approximation in the near field lends itself to various
mathematical manipulations which lead to the following suggestive
results. First, the scalar bound is a local stationary point with
respect to perturbations of the T-operators of each body, away from
the perfect isolated absorption condition. Second, $\Phi_\mathrm{sc}$
can be shown to be domain-monotonic, meaning that for fixed
$\zeta_{p}$ for each body $p \in \{\mathrm{A}, \mathrm{B}\}$, the
near-field RHT spectrum will always increase with increasing object
volumes. Together, these findings suggest that the role of
nanostructuring in enhancing near-field RHT will be limited, and this
has important implications for the theoretical and experimental design
of devices for cooling, heat dissipation, and energy generation.

\section*{Acknowledgments}

The authors would like to thank Riccardo Messina and Pengning Chao for
helpful discussions. This work was supported by the National Science
Foundation under Grants No. DMR-1454836, DMR 1420541, DGE 1148900, the
Cornell Center for Materials Research MRSEC (award no. DMR-1719875),
and the Defense Advanced Research Projects Agency (DARPA) under
agreement HR00111820046. The views, opinions and/or findings expressed
are those of the authors and should not be interpreted as representing
the official views or policies of the Department of Defense or the
U.S. Government.

\appendix

\section{Notation}

We briefly discuss the notation used through the main text and the
appendices. A vector field $\vec{v}(\vec{x})$ will be denoted as
$\ket{\vec{v}}$. The conjugated inner product is $\bracket{\vec{u},
  \vec{v}} = \int~\mathrm{d}^{3} x~\vec{u}^{\star} (\vec{x}) \cdot
\vec{v}(\vec{x})$. An operator $\mathbb{A}(\vec{x}, \vec{x}')$ will be
denoted as $\mathbb{A}$, with $\int~\mathrm{d}^{3}
x'~\mathbb{A}(\vec{x}, \vec{x}') \cdot \vec{v}(\vec{x}')$ denoted as
$\mathbb{A}\ket{\vec{v}}$. The Hermitian conjugate
$\mathbb{A}^{\dagger}$ is defined such that $\bracket{\vec{u},
  \mathbb{A}^{\dagger} \vec{v}} = \bracket{\mathbb{A} \vec{u},
  \vec{v}}$. The anti-Hermitian part of a square operator (whose
domain and range are the same size) is defined as the operator
$\asym(\mathbb{A}) = (\mathbb{A} -
\mathbb{A}^{\dagger})/(2\im)$. Finally, the trace of an operator is
$\tr(\mathbb{A}) = \int~\mathrm{d}^{3} x~\tr(\mathbb{A}(\vec{x},
\vec{x}))$. Through this paper, unless stated explicitly otherwise,
all quantities implicitly depend on $\omega$, and such dependence will
be notationally suppressed for brevity. % we also define
% $\frac{1}{\lambda_{\sigma}} = \matsigma$ for notational convenience.

\section{Proof of trace maximization via shared singular vectors}

In this section, we prove the following lemma: if operators
$\mathbb{A}_{n}$ for $n \in \{1, 2, \ldots, N\}$ have fixed singular
values labeled $\sigma^{(n)}_{i}$, then the singular vectors that
maximize $\tr[\mathbb{A}_{1} \mathbb{A}_{2} \ldots \mathbb{A}_{N}]$
are common between operators multiplied together. That is, the
singular value decomposition of $\mathbb{A}_{n}$ should follow
$\mathbb{A}_{n} = \sum_{i} \sigma^{(n)}_{i} \ket{\vec{a}^{(n)}_{i}}
\bra{\vec{a}^{(n + 1)}_{i}}$ for $n \in \{1, 2, \ldots, N - 1\}$, with
$\mathbb{A}_{N} = \sum_{i} \sigma^{(N)}_{i} \ket{\vec{a}^{(N)}_{i}}
\bra{\vec{a}^{(1)}_{i}}$, where the vectors $\ket{\vec{a}^{(n)}_{i}}$
are orthonormal for each $n$ such that
$\bracket{\vec{a}^{(n)}_{i}, \vec{a}^{(n)}_{j}} = \delta_{ij}$. This
lemma will hold even if each $\mathbb{A}_{n}$ is not square, as long
as $\mathbb{A}_{n} \mathbb{A}_{n + 1}$ forms a valid nontrivial
operator product, as these can be embedded in larger spaces padded
with more vanishing singular values. Thus, we restrict our
consideration to square operators. Moreover, associativity means
$\mathbb{A}_{n} \mathbb{A}_{n + 1} \mathbb{A}_{n + 2} =
(\mathbb{A}_{n} \mathbb{A}_{n + 1}) \mathbb{A}_{n + 2}$, and the trace
of a product of operators is invariant under cyclic permutations, so
we ultimately only consider maximizing the trace of a product of two
operators, as maximization of the trace of products of more than two
operators follows inductively from this.

To maximize $\tr[\mathbb{A}\mathbb{B}]$, we start by writing
\begin{align}
  \mathbb{A} &= \sum_{i = 1}^{N} \sigma_{i} \ket{\vec{u}_{i}} \bra{\vec{v}_{i}} \\
  \mathbb{B} &= \sum_{j = 1}^{N} \tau_{j} \ket{\vec{w}_{j}} \bra{\vec{y}_{j}}
\end{align}
where $N$ is the size of the space; this may be larger than the rank
of either $\mathbb{A}$ or $\mathbb{B}$, but the point is moot because
the singular values are fixed, whether they vanish or not, and it has
already been assumed that $\mathbb{A}$ and $\mathbb{B}$ are square. We
also assume that the singular values are ordered such that
$\sigma_{i} \geq \sigma_{i + 1}$ for all
$i \in \{1, 2, \ldots, N - 1\}$ and $\tau_{j} \geq \tau_{j + 1}$ for
all $j \in \{1, 2, \ldots, N - 1\}$. This allows for writing
\begin{equation}
  \tr[\mathbb{A}\mathbb{B}] = \sum_{i = 1}^{N} \sum_{j = 1}^{N} \sigma_{i} p_{ij} \tau_{j} q_{ji}
\end{equation}
in terms of $p_{ij} = \bracket{\vec{v}_{i}, \vec{w}_{j}}$ and
$q_{ji} = \bracket{\vec{y}_{j}, \vec{u}_{i}}$. As the singular vectors
are orthonormal, then $p_{ij}$ and $q_{ji}$ are the elements of
unitary matrices, satisfying
$\sum_{j = 1}^{N} |p_{ij}|^{2} = \sum_{i = 1}^{N} |p_{ij}|^{2} =
\sum_{j = 1}^{N} |q_{ji}|^{2} = \sum_{i = 1}^{N} |q_{ji}|^{2} = 1$. We
consider only real nonnegative $\tr[\mathbb{A}\mathbb{B}]$, so that is
maximized when $\sigma_{i} p_{ij} \tau_{j} q_{ji}$ are all
nonnegative; this means the singular vectors can be chosen without
loss of generality such that $p_{ij}$ and $q_{ji}$ are real and
nonnegative, implying $p_{ij}$ and $q_{ji}$ are the elements of
real-valued orthogonal matrices.

We use induction to prove that maximizing the trace requires that
$\{\ket{\vec{u}_{i}}\}$ be the duals of $\{\bra{\vec{y}_{j}}\}$, and
that $\{\bra{\vec{v}_{i}}\}$ by the duals of
$\{\ket{\vec{w}_{j}}\}$. The case $N = 1$ is trivial, as all
quantities are scalars. For $N = 2$, we use orthogonality to note that
$p_{1,2} = p_{2,1}$, $q_{1,2} = q_{2,1}$,
$p_{1,1} = p_{2,2} = \sqrt{1 - p_{1,2}^{2}}$, and
$q_{1,1} = q_{2,2} = \sqrt{1 - q_{1,2}^{2}}$. As a result, we may
write
$\tr[\mathbb{A}\mathbb{B}] = \sqrt{(1 - p_{1,2}^{2})(1 - q_{1,2}^{2})}
(\sigma_{1} \tau_{1} + \sigma_{2} \tau_{2}) + p_{1,2} q_{1,2}
(\sigma_{1} \tau_{2} + \sigma_{2} \tau_{1})$. As the first term in
parentheses is larger than the second term in parentheses by the
nonnegative value $(\sigma_{1} - \sigma_{2})(\tau_{1} - \tau_{2})$
given the ordering of singular values, having the left singular
vectors of one operator not be duals of the right singular vectors of
the other and vice versa could only increase the trace if
$\sqrt{(1 - p_{1,2}^{2})(1 - q_{1,2}^{2})} + p_{1,2} q_{1,2} > 1$, but
this leads to the impossible condition $0 > (p_{1,2} - q_{1,2})^{2}$,
so we can only have $p_{1,2} = q_{1,2}$ for the trace to be maximized,
implying the duality result must hold.

The inductive step assumes an arbitrary $N - 1$ and moves from there
to proving the statement for $N$. Without loss of generality, we
consider first the contribution of the largest singular value
$\tau_{1}$ of $\mathbb{B}$, namely
$\tau_{1} \ket{\vec{w}_{1}} \bra{\vec{y}_{1}}$, interacting with
$\mathbb{A} = \sum_{i} \sigma_{i} \ket{\vec{u}_{i}} \bra{\vec{v}_{i}}$
in the trace. This yields the contribution
$\sum_{i} (p_{i,1} - q_{1,i})^{2} = 2\left(1 - \sum_{i} p_{i,1}
  q_{1,i}\right) \geq 0$ using the fact that
$\sum_{i} p_{i,1}^{2} = \sum_{i} q_{1,i}^{2} = 1$, so this in turn
gives the condition $\sum_{i} p_{i,1} q_{1,i} \leq 1$. The trace can
be seen to be maximal when the above condition is saturated, so
$\sum_{i} p_{i,1} q_{1,i} = 1$, which implies $p_{i,1} = q_{1,i}$ for
every $i$. As this also holds when the roles of $\mathbb{A}$ and
$\mathbb{B}$ are interchanged, and as this can be progressively
carried out for each successively smaller singular value given
orthonormality of the singular vectors, then the duality condition
must hold, completing the proof.

\section{Derivation of radiative heat transfer formulas}

In this section, we derive the formula for the radiative heat transfer
spectrum between two bodies, valid even without assumptions about
retardation, homogeneity, locality, or isotropy. The formula depends
on individual T-operators and the vacuum Green's function, and follows
a previous derivation~\cite{VenkataramPRL2018} which considered energy
transfer by fluctuating volume currents. Through further derivation,
we also equate this formula to another formula involving the
susceptibilities and the full Maxwell Green's function, and use that
to recast the heat transfer spectrum in a Landauer form, whence we
prove that the singular values of the Landauer transmission operator
for RHT do not exceed $1/4$. Finally, we prove that the formula for
thermal emission of a single body in isolation can be derived from the
formula for RHT between two bodies in vacuum, by taking the second
body to fully enclose the first and to be perfectly absorbing, thus
taking on the role of a perfectly absorbing medium (vacuum).

\subsection{T-operator formula}

Our derivation of the heat transfer spectrum from the
fluctuation--dissipation theorem for dipole sources in each body
follows prior work~\cite{VenkataramPRL2018}, which we reproduce here
for clarity. Consider two bodies A and B in vacuum with general
susceptibilities $\VV_{p}$ for
$p \in \{\mathrm{A}, \mathrm{B}\}$ which may be inhomogeneous,
nonlocal, or anisotropic. Maxwell's equations may be written in
integral form as
\begin{align}
  \ket{\vec{E}} &= \Gvac \ket{\vec{P}} \\
  \ket{\vec{P}} &= \ket{\vec{P}^{(0)}} + \VV \ket{\vec{E}}
\end{align}
for the fields $\ket{\vec{E}}$ and total polarizations
$\ket{\vec{P}}$ in terms of the polarization sources
$\ket{\vec{P}^{(0)}}$, after defining
\begin{align}
  \ket{\vec{E}} &= \begin{bmatrix}
    \ket{\vec{E}_{\mathrm{A}}} \\
    \ket{\vec{E}_{\mathrm{B}}}
  \end{bmatrix},\,\,\, %\label{eq:blockE} \\
  \ket{\vec{P}} = \begin{bmatrix}
    \ket{\vec{P}_{\mathrm{A}}} \\
    \ket{\vec{P}_{\mathrm{B}}}
  \end{bmatrix} \label{eq:blockP} \\
%  \ket{\vec{P}^{(0)}} &= \begin{bmatrix}
%    \ket{\vec{P}^{(0)}_{\mathrm{A}}} \\
%    \ket{\vec{P}^{(0)}_{\mathrm{B}}}
%  \end{bmatrix} \label{eq:blockP0} \\
  \Gvac &= \begin{bmatrix}
    \Gvac_{\mathrm{AA}} & \Gvac_{\mathrm{AB}} \\
    \Gvac_{\mathrm{BA}} & \Gvac_{\mathrm{BB}}
  \end{bmatrix},\,\,\, %\label{eq:blockGvac} \\
  \VV^{-1} = \begin{bmatrix}
    \VV_{\mathrm{A}}^{-1} & 0 \\
    0 & \VV_{\mathrm{B}}^{-1}
  \end{bmatrix} \label{eq:blockVinv}
\end{align}  
in block form for the material degrees of freedom constituting each
object. By defining the total T-operator via
\begin{equation} \label{eq:blockTinv}
  \TT^{-1} = \begin{bmatrix}
    \TT_{\mathrm{A}}^{-1} & -\Gvac_{\mathrm{AB}} \\
    -\Gvac_{\mathrm{BA}} & \TT_{\mathrm{B}}^{-1}
  \end{bmatrix}
\end{equation}
where $\TT_{p}^{-1} = \VV_{p}^{-1} - \Gvac_{pp}$,
then Maxwell's equations can be formally solved to yield
\begin{equation} \label{eq:Maxwellsolve}
  \begin{split}
    \ket{\vec{E}} &= \Gvac\TT\VV^{-1} \ket{\vec{P}^{(0)}} \\
    \ket{\vec{P}} &= \TT\VV^{-1} \ket{\vec{P}^{(0)}}
  \end{split}
\end{equation}
obtained by applying formulas for the block matrix inverse to compute
$\TT$. We also define the projection operators,
\begin{equation}
  \begin{split}
    \II_{\mathrm{A}} &= \begin{bmatrix}
      \II_{\mathrm{A}} & 0 \\
      0 & 0
    \end{bmatrix}, \,\,\, %\\
    \II_{\mathrm{B}} = \begin{bmatrix}
      0 & 0 \\
      0 & \II_{\mathrm{B}}
    \end{bmatrix},
  \end{split}
\end{equation}
such that (abusing notation) $\II_{p}$ is the projection onto the
material degrees of freedom of body $p$.

We consider the energy flow from fluctuating dipole sources only in
body A into material degrees of freedom in body B, noting that
reciprocity would yield the same heat transfer if the roles of bodies
A and B were interchanged. This means
%\begin{equation*}
$
  \ket{\vec{P}^{(0)}} = \begin{bmatrix}
    \ket{\vec{P}^{(0)}_{\mathrm{A}}} \\
    0
  \end{bmatrix}  
$
%\end{equation*}
defines the fluctuating sources in body A. The heat transfer spectrum
is the ensemble-averaged work, denoted by $\bracket{\cdots}$, done by
the field,
\begin{equation}
  \Phi = \frac{1}{2}\Re(\bracket{\bracket{\mathbb{I}_{\mathrm{B}} \vec{E}, \mathbb{I}_{\mathrm{B}} \vec{J}}})
\end{equation}
where $\ket{\vec{J}} = -\im\omega\ket{\vec{P}}$. Using the Hermiticity
and idempotence of $\mathbb{I}_{p}$ yields
$\Phi = -\frac{\omega}{4\im}
(\bracket{\bracket{\mathbb{I}_{\mathrm{B}} \vec{P}, \vec{E}}} -
\bracket{\bracket{\vec{E}, \mathbb{I}_{\mathrm{B}} \vec{P}}})$, and
using the results of~\eqref{Maxwellsolve} gives
\begin{equation}
  \Phi = -\frac{\omega}{2} \bracket{\bracket{\vec{P}^{(0)}_{\mathrm{A}}, \II_{\mathrm{A}} \VV^{-1\dagger} \TT^{\dagger}~\asym(\II_{\mathrm{B}} \Gvac) \TT \VV^{-1} \II_{\mathrm{A}} \vec{P}^{(0)}_{\mathrm{A}}}}
\end{equation}
in terms of the fluctuating sources
$\ket{\vec{P}^{(0)}_{\mathrm{A}}}$. As these fluctuations are thermal
in nature, their correlations are given by the
fluctuation--dissipation theorem
\begin{equation}
  \bracket{\ket{\vec{P}^{(0)}_{\mathrm{A}}} \bra{\vec{P}^{(0)}_{\mathrm{A}}}} = \frac{4}{\pi\omega} \asym(\mathbb{V}_{\mathrm{A}})
\end{equation}
(suppressing the Planck function $\Pi$ as it has already been factored
to be separate from $\Phi$), yielding
\begin{equation}
  \Phi = -\frac{2}{\pi} \tr(\asym(\VV_{\mathrm{A}}^{-1\dagger})
  \II_{\mathrm{A}} \TT^{\dagger}~\asym(\II_{\mathrm{B}} \Gvac)
  \TT\II_{\mathrm{A}})
\end{equation}
as the dressed radiative heat transfer spectrum.

To prove equivalence of this expression for $\Phi$ to that involving
only $\Gvac$ and $\TT_{p}$, it is useful to explicitly invoke
reciprocity: $\VV_{p}^{\top} = \VV_{p}$, $\TT_{p}^{\top} = \TT_{p}$,
and $(\Gvac_{pq})^{\top} = \Gvac_{qp}$ for
$p, q \in \{\mathrm{A}, \mathrm{B}\}$, so this also means
$\VV_{p}^{\dagger} = \VV_{p}^{\star}$,
$\TT_{p}^{\dagger} = \TT_{p}^{\star}$, and
$(\Gvac_{pq})^{\dagger} = \mathbb{G}^{\mathrm{vac}\star}_{qp}$; this
means $\asym(\mathbb{A}) = \Im(\mathbb{A})$ for
$\mathbb{A} \in \{\VV_{p}, \TT_{p}, \Gvac_{pp}\}$. This allows for
writing the operators
\begin{align*}
  \asym(\II_{\mathrm{B}} \Gvac) &= \begin{bmatrix}
    0 & -\mathbb{G}^{\mathrm{vac}\star}_{\mathrm{AB}}/(2\im) \\
    \Gvac_{\mathrm{BA}}/(2\im) & \Im(\Gvac_{\mathrm{BB}})
  \end{bmatrix} \\
  \TT\II_{\mathrm{A}} &= \begin{bmatrix}
    (\TT_{\mathrm{A}}^{-1} - \Gvac_{\mathrm{AB}} \TT_{\mathrm{B}}
    \Gvac_{\mathrm{BA}})^{-1} \\
    \TT_{\mathrm{B}} \Gvac_{\mathrm{BA}} (\TT_{\mathrm{A}}^{-1} -
    \Gvac_{\mathrm{AB}} \TT_{\mathrm{B}} \Gvac_{\mathrm{BA}})^{-1}
  \end{bmatrix} \\
  \II_{\mathrm{A}}\TT^{\dagger} &= \begin{bmatrix}
    (\TT_{\mathrm{A}}^{-1\star} -
    \GG^{\mathrm{vac}\star}_{\mathrm{AB}}
    \TT^{\star}_{\mathrm{B}}
    \GG^{\mathrm{vac}\star}_{\mathrm{BA}})^{-1} &
\\
    (\TT_{\mathrm{A}}^{-1\star} -
    \GG^{\mathrm{vac}\star}_{\mathrm{AB}}
    \TT^{\star}_{\mathrm{B}}
    \GG^{\mathrm{vac}\star}_{\mathrm{BA}})^{-1}
    \GG^{\mathrm{vac}\star}_{\mathrm{AB}}
    \TT^{\star}_{\mathrm{B}} 
  \end{bmatrix}
\end{align*}
in block matrix form, where the projection onto A allows for
truncation to the appropriate block column or row for notational
convenience; note that $\II_{\mathrm{A}}\TT^{\dagger}$ should actually
be a row vector, but has been written as a column for ease of
reading. Multiplying these matrices together, it can be noted that
$\Gvac_{\mathrm{BA}} (\TT_{\mathrm{A}}^{-1} - \Gvac_{\mathrm{AB}}
\TT_{\mathrm{B}} \Gvac_{\mathrm{BA}})^{-1} = \Gvac_{\mathrm{BA}}
(\II_{\mathrm{A}} - \TT_{\mathrm{A}} \Gvac_{\mathrm{AB}}
\TT_{\mathrm{B}} \Gvac_{\mathrm{BA}})^{-1} \TT_{\mathrm{A}} =
(\II_{\mathrm{B}} - \Gvac_{\mathrm{BA}} \TT_{\mathrm{A}}
\Gvac_{\mathrm{AB}} \TT_{\mathrm{B}})^{-1} \Gvac_{\mathrm{BA}}
\TT_{\mathrm{A}} = \Ss^{\mathrm{A}}_{\mathrm{B}} \Gvac_{\mathrm{BA}}
\TT_{\mathrm{A}}$ using the definition of the scattering operator
$\Ss^{\mathrm{A}}_{\mathrm{B}} = (\II_{\mathrm{B}} - \Gvac_{\mathrm{BA}}
\TT_{\mathrm{A}} \Gvac_{\mathrm{AB}}
\TT_{\mathrm{B}})^{-1}$. Additionally, using the definition
$\TT_{p}^{-1} = \VV_{p}^{-1} - \Gvac_{pp}$, it is easy to prove that
$\Im(\TT_{\mathrm{B}}) - \TT_{\mathrm{B}}^{\star}
\Im(\Gvac_{\mathrm{BB}}) \TT_{\mathrm{B}} = \TT_{\mathrm{B}}^{\star}
\Im(\VV_{\mathrm{B}}^{-1\star}) \TT_{\mathrm{B}}$, and likewise
$\Im(\TT_{\mathrm{A}}) - \TT_{\mathrm{A}} \Im(\Gvac_{\mathrm{AA}})
\TT_{\mathrm{A}}^{\star} = \TT_{\mathrm{A}}
\Im(\VV_{\mathrm{A}}^{-1\star}) \TT_{\mathrm{A}}^{\star}$. Thus, the
result is~\eqref{exactPhibothT} in the main text, as expected.

\subsection{Derivation of Green's function heat transfer formula}

Our derivation of the bounds in the main text relies on the
relationship between the heat transfer spectrum $\Phi$ written in
terms of the vacuum Green's function and the T-operators of individual
objects, to the heat transfer formula~\cite{JinPRB2019}
\begin{equation}
  \Phi = \frac{2}{\pi} \tr[\asym(\VV_{\mathrm{A}}) \GG_{\mathrm{BA}}^{\dagger} \asym(\VV_{\mathrm{B}}) \GG_{\mathrm{BA}}]
\end{equation}
where
$\mathbb{G} = (\mathbb{G}^{\mathrm{vac}-1} - \VV_{\mathrm{A}} -
\VV_{\mathrm{B}})^{-1}$ is the full Maxwell Green's function in
the presence of both bodies, with the block $\GG_{\mathrm{BA}}$
representing the fields in body B due to dipole sources in body A. We
start with the T-operator by rewriting
\begin{multline}
%\begin{equation}
  \Phi(\omega) = \frac{2}{\pi} \tr[\asym(\VV_{\mathrm{A}})
  \VV_{\mathrm{A}}^{-1\dagger} \TT_{\mathrm{A}}^{\dagger}
  (\Ss^{\mathrm{A}}_{\mathrm{B}} \Gvac_{\mathrm{BA}})^{\dagger}
  \TT_{\mathrm{B}}^{\dagger} \times
  \\
  \VV_{\mathrm{B}}^{-1\dagger} \asym(\VV_{\mathrm{B}})
  \VV_{\mathrm{B}}^{-1} \TT_{\mathrm{B}} \Ss^{\mathrm{A}}_{\mathrm{B}}
  \Gvac_{\mathrm{BA}} \TT_{\mathrm{A}} \VV_{\mathrm{A}}^{-1}]
%\end{equation}
\end{multline}
after using
$\asym(\VV_{\mathrm{B}}^{-1\dagger}) = \VV_{\mathrm{B}}^{-1\dagger}
\asym(\VV_{\mathrm{B}}) \VV_{\mathrm{B}}^{-1}$ and
$\asym(\VV_{\mathrm{A}}^{-1\dagger}) = \VV_{\mathrm{A}}^{-1}
\asym(\VV_{\mathrm{A}}) \VV_{\mathrm{A}}^{-1\dagger}$ along with
invariance of the trace under cyclic permutations of operator
products. From this, it can be seen that the two expressions for
$\Phi(\omega)$ are guaranteed to be the same if the operator
$\GG_{\mathrm{BA}}$ is the same as
$\VV_{\mathrm{B}}^{-1} \TT_{\mathrm{B}} \Ss^{\mathrm{A}}_{\mathrm{B}}
\Gvac_{\mathrm{BA}} \TT_{\mathrm{A}} \VV_{\mathrm{A}}^{-1}$. We use
the fact that
$\VV_{\sigma}^{-1} = \TT_{\sigma}^{-1} + \Gvac_{\sigma\sigma}$ to say
that
\begin{equation*}
  \VV_{\mathrm{B}}^{-1} \TT_{\mathrm{B}}
  \Ss^{\mathrm{A}}_{\mathrm{B}} \Gvac_{\mathrm{BA}} \TT_{\mathrm{A}}
  \VV_{\mathrm{A}}^{-1} = (\II_{\mathrm{B}} + \Gvac_{\mathrm{BB}} \TT_{\mathrm{B}}) \Ss^{\mathrm{A}}_{\mathrm{B}} \Gvac_{\mathrm{BA}} (\II_{\mathrm{A}} + \TT_{\mathrm{A}} \mathrm{G}^{\mathrm{vac}}_{\mathrm{AA}})
\end{equation*}
must hold. To prove that this is equal to $\GG_{\mathrm{BA}}$,
we use the definition
\begin{equation}
  \mathbb{G} = \Gvac + \Gvac \TT \Gvac
\end{equation}
in conjunction with definitions of $\Gvac$ and
$\TT$ as $2\times 2$ block matrices in~\eqref{blockTinv} to
write
\begin{equation}
  \GG_{\mathrm{BA}} = \Gvac_{\mathrm{BA}} + \begin{bmatrix}
    \Gvac_{\mathrm{BA}} & \Gvac_{\mathrm{BB}}
  \end{bmatrix} \TT \begin{bmatrix}
    \Gvac_{\mathrm{AA}} \\
    \Gvac_{\mathrm{BA}} &
  \end{bmatrix}
\end{equation}
for this system. Performing this matrix multiplication, recognizing
that
$\Gvac_{\mathrm{BA}} \TT_{\mathrm{A}} \Gvac_{\mathrm{AB}}
(\TT_{\mathrm{B}}^{-1} - \Gvac_{\mathrm{BA}} \TT_{\mathrm{A}}
\Gvac_{\mathrm{AB}})^{-1} = \Ss^{\mathrm{A}}_{\mathrm{B}} - \II_{\mathrm{B}}$,
using the fact definition of
$\Ss^{\mathrm{A}}_{\mathrm{B}}%  = (\II_{\mathrm{B}} - \Gvac_{\mathrm{BA}}
% \TT_{\mathrm{A}} \Gvac_{\mathrm{AB}} \TT_{\mathrm{B}})^{-1}
$, and
collecting and canceling terms leads to the proof of the equality
$\GG_{\mathrm{BA}} = \VV_{\mathrm{B}}^{-1} \TT_{\mathrm{B}}
\Ss^{\mathrm{A}}_{\mathrm{B}} \Gvac_{\mathrm{BA}} \TT_{\mathrm{A}}
\VV_{\mathrm{A}}^{-1}$.

\subsection{Landauer bounds on heat transfer singular values}

We now prove that radiative heat transfer between arbitrarily shaped
bodies can also be expressed as the trace of a transmission matrix
whose singular values can be bounded above, similar to previously
derived bounds in planar media. This relation intuitively connects the
finite value of the RHT bounds and approximate low rank of
$\GG_{\mathrm{BA}}$, and can be proved as follows. For this, we use
the cyclic property of the trace to define
\begin{equation}
  \Phi = \frac{2}{\pi} \tr(\mathbb{Q}^{\dagger} \mathbb{Q})
\end{equation}
where
$\mathbb{Q} = \Im(\VV_{\mathrm{B}})^{1/2} \GG_{\mathrm{BA}}
\Im(\VV_{\mathrm{A}})^{1/2}$ is the heat transmission operator.

The definition
$\mathbb{G}^{-1} = \mathbb{G}^{\mathrm{vac}-1} - (\VV_{\mathrm{A}} +
\VV_{\mathrm{B}})$ along with the fact that the vacuum Maxwell
operator $\mathbb{G}^{\mathrm{vac}-1}$ is real-valued in position
space allows for writing
\begin{equation}
  \asym(\mathbb{G}) = \mathbb{G}^{\dagger} \asym(\VV_{\mathrm{A}} + \VV_{\mathrm{B}}) \mathbb{G}
\end{equation}
which relates dissipation in polarization currents and electromagnetic
fields in equilibrium. Additionally, the fact that $\asym(\VV_{p})$ is
a Hermitian positive-definite operator for each body
$p \in \{\mathrm{A}, \mathrm{B}\}$ means it has a unique square root
$\asym(\VV_{\mathrm{p}})^{1/2}$. Rearranging the above equation,
multiplying both sides by $2~\asym(\VV_{\mathrm{A}})^{1/2}$, and
adding $\II_{\mathrm{A}}$ to both sides gives
\begin{multline*}
  4~\asym(\VV_{\mathrm{A}})^{1/2} \mathbb{G}^{\dagger} \asym(\VV_{\mathrm{B}}) \mathbb{G} \asym(\VV_{\mathrm{A}})^{1/2} 
\\ 
+ 4~\asym(\VV_{\mathrm{A}})^{1/2} \mathbb{G}^{\dagger} \asym(\VV_{\mathrm{A}}) \mathbb{G} \asym(\VV_{\mathrm{A}})^{1/2} \\ + 2\im (\asym(\VV_{\mathrm{A}})^{1/2} \mathbb{G} \asym(\VV_{\mathrm{A}})^{1/2} 
\\ 
- \asym(\VV_{\mathrm{A}})^{1/2} \mathbb{G}^{\dagger} \asym(\VV_{\mathrm{A}})^{1/2}) + \II_{\mathrm{A}} = \II_{\mathrm{A}}
\end{multline*}
where we recognize the equality
$\mathbb{Q}^{\dagger} \mathbb{Q} = \asym(\VV_{\mathrm{A}})^{1/2}
\GG^{\dagger} \asym(\VV_{\mathrm{B}}) \GG
\asym(\VV_{\mathrm{A}})^{1/2}$. Following this substitution, this may
be factored as
\begin{multline}
  4~\mathbb{Q}^{\dagger} \mathbb{Q} + \left(\II_{\mathrm{A}} +
    2\im~\asym(\VV_{\mathrm{A}})^{1/2} \GG_{\mathrm{AA}}
    \asym(\VV_{\mathrm{A}})^{1/2} \right)^{\dagger}
  \times \\
  \left(\II_{\mathrm{A}} + 2\im~\asym(\VV_{\mathrm{A}})^{1/2}
    \GG_{\mathrm{AA}} \asym(\VV_{\mathrm{A}})^{1/2} \right) =
  \II_{\mathrm{A}}
\end{multline}
where $\GG$ has been replaced by its blocks $\GG_{\mathrm{AA}}$ and
$\GG_{\mathrm{AB}}$ due to multiplications on each each side by
$\asym(\VV_{p})^{1/2}$ for $p \in \{\mathrm{A}, \mathrm{B}\}$ (and
likewise for $\mathbb{G}^{\dagger}$). This expression is the sum of
two Hermitian positive-semidefinite operators equal to the identity;
though this has been done for body A, reciprocity of heat transfer
yields a similar expression in terms of the operators for body
B. Consequently, the singular values of the operator
$\mathbb{Q}^{\dagger} \mathbb{Q}$ entering the trace expression for
$\Phi(\omega)$ must all be less than or equal to $1/4$. We emphasize
that this derivation is valid for compact or extended structures of
\emph{arbitrary} geometry, without any need to expand heat transfer in
terms of incoming and outgoing plane waves specific to translationally
symmetric systems~\cite{PendryJPCM1999}.

\subsection{Single-body thermal radiation from two-body radiative heat
  transfer}

In this section, we prove that the formula for thermal emission of a
single body in isolation in vacuum can be derived by starting from the
formula for heat transfer between two bodies in vacuum under the
following conditions. We take body A to be the thermal emitter in
question, while body B is taken to fully surround body A as a shell of
inner radius $r_{\mathrm{B}}$ and outer radius $R_{\mathrm{B}}$ with
susceptibility
$\VV_{\mathrm{B}} = \chi_{\mathrm{B}} \II_{\mathrm{B}}$, and take the
simultaneous limits $\Im(\chi_{\mathrm{B}}) \to 0$ and
$\omega r_{\mathrm{B}} / c \to \infty$ constrained by
$\omega (R_{\mathrm{B}} - r_{\mathrm{B}}) / c \to \infty$ and
$\omega (R_{\mathrm{B}} - r_{\mathrm{B}}) \Im(\chi_{\mathrm{B}}) / c
\to 1$, in which case body B takes on the role of a perfectly
absorbing medium.

To start, we note that
$\TT_{\mathrm{B}}^{-1} = \VV_{\mathrm{B}}^{-1} - \Gvac_{\mathrm{BB}}$
in conjunction with reciprocity allows for
writing~\eqref{exactPhibothT} as
\begin{multline*}
  \Phi = \frac{2}{\pi} \tr[\TT_{\mathrm{B}}
  \Ss^{\mathrm{A}}_{\mathrm{B}} \asym(\VV_{\mathrm{B}}^{-1\dagger})
  \Ss^{\mathrm{A}\dagger}_{\mathrm{B}} \TT_{\mathrm{B}}^{\dagger}
  \times \\ (\Gvac_{\mathrm{AB}})^{\dagger} (\asym(\TT_{\mathrm{A}}) -
  \TT_{\mathrm{A}}^{\dagger} \asym(\Gvac_{\mathrm{AA}})
  \TT_{\mathrm{A}}) \Gvac_{\mathrm{AB}}].
\end{multline*}
In the aforementioned size and susceptibility limits for body B,
$\TT_{\mathrm{B}} \to 0$ so
$\Ss^{\mathrm{A}}_{\mathrm{B}} \to \II_{\mathrm{B}}$, and
$\TT_{\mathrm{B}} \Ss^{\mathrm{A}}_{\mathrm{B}}
\asym(\VV_{\mathrm{B}}^{-1\dagger})
\Ss^{\mathrm{A}\dagger}_{\mathrm{B}} \TT_{\mathrm{B}}^{\dagger} \to
\asym(\VV_{\mathrm{B}})$. Using reciprocity, this yields
$\Phi = \frac{2}{\pi} \tr[\Gvac_{\mathrm{AB}} \Im(\VV_{\mathrm{B}})
\GG_{\mathrm{BA}}^{\mathrm{vac}\star} (\Im(\TT_{\mathrm{A}}) -
\TT_{\mathrm{A}}^{\star} \Im(\Gvac_{\mathrm{AA}})
\TT_{\mathrm{A}})$. For a system with a general susceptibility $\VV$
and Maxwell Green's function
$\GG = (\GG^{\mathrm{vac}-1} - \VV)^{-1}$, the relations
$\GG \Im(\VV) \GG^{\star} = \GG^{\star} \Im(\VV) \GG = \Im(\GG)$ will
always hold. Considering B in isolation, the simultaneous constrained
limits of infinite size and infinitesimal susceptibility mean that
$\Gvac_{\mathrm{AB}} \Im(\VV_{\mathrm{B}})
\GG_{\mathrm{BA}}^{\mathrm{vac}\star} \to
\Im(\Gvac_{\mathrm{AA}})$. Finally, this yields the emission formula
\begin{equation}
  \Phi = \frac{2}{\pi} \tr[\Im(\Gvac)(\Im(\TT) - \TT^{\star} \Im(\Gvac) \TT)]
\end{equation}
in agreement with the formula derived by Kr\"{u}ger et
al~\cite{KrugerPRB2012}, where the subscripts A have been dropped as
there is only one material body under consideration given that body B
has effectively vanished.

\section{Nonnegative far-field scattering}

In this section, we derive conditions on operators describing the
response of bodies $p \in \{\mathrm{A}, \mathrm{B}\}$ for far-field
scattering to be nonnegative. In general, given a susceptibility $\VV$
and an associated T-operator $\TT$, the far-field scattered power from
a given incident field $\ket{\vec{E}^{\mathrm{inc}}}$ is
$\frac{\omega}{2} \bracket{\vec{E}^{\mathrm{inc}}, (\Im(\TT) -
  \TT^{\star} \Im(\VV^{-1\star})\TT)\vec{E}^{\mathrm{inc}}}$, and for
this to be nonnegative, the operator
$\Im(\TT) - \TT^{\star} \Im(\VV^{-1\star})\TT$ must be
positive-semidefinite. This must hold true for each body in isolation,
meaning $\Im(\TT_{p}) - \TT_{p}^{\star} \Im(\VV_{p}^{-1\star})\TT_{p}$
must be positive-semidefinite for each
$p \in \{\mathrm{A}, \mathrm{B}\}$. However, more conditions can be
derived when the two bodies are proximate to each other.

As a first step, we show that the operator
$\YY_{\mathrm{B}} = (\TT_{\mathrm{B}}^{-1} - \Gvac_{\mathrm{BA}}
\TT_{\mathrm{A}} \Gvac_{\mathrm{AB}})^{-1} = \TT_{\mathrm{B}}
\Ss^{\mathrm{A}}_{\mathrm{B}}$ is an effective T-operator for body B dressed by the
proximity of body A. In particular, the definitions~\eqref{blockVinv}
and~\eqref{blockTinv} can be plugged into~\eqref{Maxwellsolve} and
rearranged in order to write
\begin{equation*}
  \ket{\vec{P}_{\mathrm{B}}} = \YY_{\mathrm{B}} (\Gvac_{\mathrm{BA}} \TT_{\mathrm{A}} \VV_{\mathrm{A}}^{-1} \ket{\vec{P}^{(0)}_{\mathrm{A}}} + \VV_{\mathrm{B}}^{-1} \ket{\vec{P}^{(0)}_{\mathrm{B}}})
\end{equation*}
and if only sources in A are relevant, then we may set
$\ket{\vec{P}^{(0)}_{\mathrm{B}}} \to 0$ and define an effective
incident field
$\ket{\vec{E}^{\mathrm{inc}}} = \Gvac_{\mathrm{BA}} \TT_{\mathrm{A}}
\VV_{\mathrm{A}}^{-1} \ket{\vec{P}^{(0)}_{\mathrm{A}}}$ which depends
on multiple scattering within A but not on any properties of B apart
from projection onto its volumetric degrees of freedom. This means
$\ket{\vec{P}_{\mathrm{B}}} = \YY_{\mathrm{B}}
\ket{\vec{E}^{\mathrm{inc}}}$, which is interpreted to mean that the
total induced polarization in B arises from the response of B dressed
in the presence of A, namely $\YY_{\mathrm{B}}$, acting on the
effective incident field $\ket{\vec{E}^{\mathrm{inc}}}$ accounting
only for body A; this is analogous to $\TT_{\mathrm{B}}$ which relates
the total polarization induced in B to incident fields in vacuum.

Given this and the fact that
$\ket{\vec{E}_{\mathrm{B}}} = \VV_{\mathrm{B}}^{-1}
\ket{\vec{P}_{\mathrm{B}}}$ after setting
$\ket{\vec{P}^{(0)}_{\mathrm{B}}} \to 0$, the scattered power only
from body B (in the presence of body A) may be written as the
difference between extinction and absorption powers only from body B
(in the presence of body A), namely
$\frac{\omega}{2} \left(\Im(\bracket{\vec{E}^{\mathrm{inc}},
    \vec{P}_{\mathrm{B}}}) - \bracket{\vec{E}_{\mathrm{B}},
    \vec{P}_{\mathrm{B}}}\right) = \frac{\omega}{2}
\bracket{\vec{E}^{\mathrm{inc}}, (\Im(\YY_{\mathrm{B}}) -
  \YY_{\mathrm{B}}^{\star} \Im(\VV_{\mathrm{B}}^{-1\star})
  \YY_{\mathrm{B}}) \vec{E}^{\mathrm{inc}}}$. Nonnegativity of this
quantity for any $\ket{\vec{P}^{(0)}_{\mathrm{A}}}$, or more generally
any $\ket{\vec{E}^{\mathrm{inc}}}$, means that
$\Im(\YY_{\mathrm{B}}) - \YY_{\mathrm{B}}^{\star}
\Im(\VV_{\mathrm{B}}^{-1\star}) \YY_{\mathrm{B}}$ must be
positive-semidefinite.

However, nonnegativity of far-field scattering from the system in
general means that upon evaluating the inverse of~\eqref{blockTinv},
the operator $\Im(\TT) - \TT^{\star} \Im(\VV^{-1\star})\TT$ must be
positive-semidefinite, which means in turn that each of its diagonal
blocks must be positive-semidefinite. Manipulating operators allows
for showing that the bottom-right block is $\Im(\YY_{\mathrm{B}}) -
\YY_{\mathrm{B}} \left(\Im(\VV_{\mathrm{B}}^{-1\star}) +
\Gvac_{\mathrm{BA}} \TT_{\mathrm{A}} \Im(\VV_{\mathrm{A}}^{-1\star})
\TT_{\mathrm{A}}^{\star} \GG_{\mathrm{AB}}^{\mathrm{vac}\star}\right)
\YY_{\mathrm{B}}^{\star}$, which involves another
positive-semidefinite operator (as $\Im(\VV_{\mathrm{A}}^{-1\star})$
is positive-semidefinite and is multiplied on its left and right by
operators which are Hermitian adjoints of each other, leaving the
positive-semidefiniteness unaffected) subtracted from the operator
$\Im(\YY_{\mathrm{B}}) - \YY_{\mathrm{B}}^{\star}
\Im(\VV_{\mathrm{B}}^{-1\star}) \YY_{\mathrm{B}}$. Therefore, the
positive-semidefiniteness of $\Im(\YY_{\mathrm{B}}) - \YY_{\mathrm{B}}
\left(\Im(\VV_{\mathrm{B}}^{-1\star}) + \Gvac_{\mathrm{BA}}
\TT_{\mathrm{A}} \Im(\VV_{\mathrm{A}}^{-1\star})
\TT_{\mathrm{A}}^{\star} \GG_{\mathrm{AB}}^{\mathrm{vac}\star}\right)
\YY_{\mathrm{B}}^{\star}$ yields stronger bounds on the singular
values of $\YY_{\mathrm{B}}$ than the positive-semidefiniteness of
$\Im(\YY_{\mathrm{B}}) - \YY_{\mathrm{B}}^{\star}
\Im(\VV_{\mathrm{B}}^{-1\star}) \YY_{\mathrm{B}}$ alone, because the
former also subtracts the absorption in body A in the presence of body
B, whereas the latter does not.

\section{Alternative derivations of scalar approximation and Born
  bound}

In this section, we provide alternative derivations of the scalar
approximation $\Phi_{\mathrm{sc}}$ and the Born bound
$\Phi_{\mathrm{Born}}$. We start by writing the RHT spectrum
in~\eqref{exactPhibothT} as
\begin{equation}
  \Phi = \frac{2}{\pi\zeta_{\mathrm{A}} \zeta_{\mathrm{B}}}
  \left\lVert \TT_{\mathrm{B}} \Ss^{\mathrm{A}}_{\mathrm{B}}
  \Gvac_{\mathrm{BA}} \TT_{\mathrm{A}} \right\rVert_{\mathrm{F}}^{2}
\end{equation}
in terms of the scattering operator $\Ss^{\mathrm{A}}_{\mathrm{B}}$,
after using the fact that $\asym(\VV_{p}^{-1\dagger}) =
\frac{1}{\zeta_{p}} \II_{p}$ for each body $p \in \{\mathrm{A},
\mathrm{B}\}$. The scalar approximation follows from choosing the
T-operator for each body to be $\TT_{p} = \im\tau_{p}\II_{p}$ for some
choice of uniform singular values $\tau_{p} \in [0, \zeta_{p}]$. This
gives $\Phi_{\mathrm{sc}} = \frac{2\tau_{\mathrm{A}}^{2}
  \tau_{\mathrm{B}}^{2}}{\pi\zeta_{\mathrm{A}} \zeta_{\mathrm{B}}}
\sum_{i} \frac{g_{i}^{2}}{(1 + \tau_{\mathrm{A}} \tau_{\mathrm{B}}
  g_{i}^{2})^{2}}$, and it can be seen that the contribution from each
channel $i$ monotonically increases with increasing $\tau_{\mathrm{A}}
\tau_{\mathrm{B}}$. Thus, the only limit comes from the constraints on
$\tau_{p}$, which therefore must saturate at $\tau_{p} =
\zeta_{p}$. This is equivalent to directly plugging in $\TT_{p} =
\im\zeta_{p} \II_{p}$, which immediately yields $\Phi_{\mathrm{sc}} =
\frac{2}{\pi} \zeta_{\mathrm{A}} \zeta_{\mathrm{B}} \left\lVert
\Ss^{\mathrm{A}}_{\mathrm{B}} \Gvac_{\mathrm{BA}}
\right\rVert_{\mathrm{F}}^{2}$ where $\Ss^{\mathrm{A}}_{\mathrm{B}} =
(\II_{\mathrm{B}} + \zeta_{\mathrm{A}} \zeta_{\mathrm{B}}
\Gvac_{\mathrm{BA}} \Gvac_{\mathrm{AB}})^{-1}$ in the scalar
approximation. Using the singular value decomposition of
$\Gvac_{\mathrm{BA}}$, particularly the facts that
$\bracket{\vec{a}^{(j)}, \vec{a}^{(j)\star}} = \bracket{\vec{b}^{(j)},
  \vec{b}^{(j)\star}} \im$ for each channel $j$, leads to the
expression in the main text in terms of the singular values $\rho_{j}$
of $\Gvac_{\mathrm{BA}}$.

The Born bound can be derived from $\Phi_{\mathrm{sc}} = \frac{2}{\pi}
\zeta_{\mathrm{A}} \zeta_{\mathrm{B}} \left\lVert
\Ss^{\mathrm{A}}_{\mathrm{B}} \Gvac_{\mathrm{BA}}
\right\rVert_{\mathrm{F}}^{2}$ by taking the scattering operator
$\Ss^{\mathrm{A}}_{\mathrm{B}} \to \II_{\mathrm{B}}$, which is exactly
a Born approximation. This exactly gives $\Phi_{\mathrm{Born}} =
\frac{2}{\pi} \zeta_{\mathrm{A}} \zeta_{\mathrm{B}} \left\lVert
\Gvac_{\mathrm{BA}} \right\rVert_{\mathrm{F}}^{2}$, and substituting
the singular values of $\Gvac_{\mathrm{BA}}$ in the Frobenius norm
squared gives the result in the main text. Alternatively, the operator
$\Ss^{\mathrm{A}}_{\mathrm{B}}$ contributes the denominators $(1 +
\zeta_{\mathrm{A}} \zeta_{\mathrm{B}} g_{i})^{2}$ to each channel $i$,
so the replacement $\Ss^{\mathrm{A}}_{\mathrm{B}} \to
\II_{\mathrm{B}}$ makes those denominators disappear. In any case, the
result in the main text is recovered by neglecting multiple
scattering between the two bodies.

\section{Proof of domain monotonicity of singular values of
  $\Gvac_{\mathrm{BA}}$}

In this section, we prove that the singular values $g_{i}$ of
$\Gvac_{\mathrm{BA}}$ are domain monotonic. The singular values of
$\Gvac_{\mathrm{BA}}$ are the eigenvalues of
$\Gvac_{\mathrm{BA}} (\Gvac_{\mathrm{BA}})^{\dagger}$. We consider the
effects of a perturbative addition of volume only to body A; a
perturbative effect on body B can be considered through reciprocity,
and the proof will remain the same. This allows for writing the block
row vector of operators
\begin{equation}
  \Gvac_{\mathrm{BA}} = \begin{bmatrix}
    \Gvac_{\mathrm{BA}_{0}} & \Gvac_{\mathrm{B}\Delta\mathrm{A}}
  \end{bmatrix}
\end{equation}
where $\Gvac_{\mathrm{BA}_{0}}$ is the operator propagating fields in
vacuum from the unperturbed volume $\mathrm{A}_{0}$ to body B, and
$\Gvac_{\mathrm{B}\Delta\mathrm{A}}$ is the operator propagating
fields in vacuum from the perturbative volume $\Delta\mathrm{A}$ to
body B. Using reciprocity, we may then write
\begin{multline}
  \Gvac_{\mathrm{BA}} (\Gvac_{\mathrm{BA}})^{\dagger} =
  \Gvac_{\mathrm{BA}_{0}} \GG^{\mathrm{vac}\star}_{\mathrm{A}_{0}
    \mathrm{B}} + \Gvac_{\mathrm{BA}_{0}}
  \GG^{\mathrm{vac}\star}_{\Delta\mathrm{AB}} +
  \Gvac_{\mathrm{B}\Delta\mathrm{A}}
  \GG^{\mathrm{vac}\star}_{\mathrm{A}_{0} \mathrm{B}} \\ +
  \Gvac_{\mathrm{B}\Delta\mathrm{A}}
  \GG^{\mathrm{vac}\star}_{\Delta\mathrm{AB}}
\end{multline}
for which the first term $\Gvac_{\mathrm{BA}_{0}}
\GG^{\mathrm{vac}\star}_{\mathrm{A}_{0} \mathrm{B}}$ is the Hermitian
positive-semidefinite unperturbed operator, the terms
$\Gvac_{\mathrm{BA}_{0}} \GG^{\mathrm{vac}\star}_{\Delta\mathrm{AB}}$
and $\Gvac_{\mathrm{B}\Delta\mathrm{A}}
\GG^{\mathrm{vac}\star}_{\mathrm{A}_{0} \mathrm{B}}$ vanish because
the projections onto the volume $\mathrm{A}_{0}$ and
$\Delta\mathrm{A}$ are orthogonal to each other, and
$\Gvac_{\mathrm{B}\Delta\mathrm{A}}
\GG^{\mathrm{vac}\star}_{\Delta\mathrm{AB}}$ is the Hermitian
positive-semidefinite perturbation. From Rayleigh-Schr\"{o}dinger
perturbation theory, if $\rho_{i}$ is an unperturbed singular value of
$\Gvac_{\mathrm{BA}_{0}}$ with $\ket{\vec{b}_{i}}$ being the
corresponding normalized right singular vector, then the perturbation
to $\rho_{i}$ is $\bracket{\vec{b}_{i},
  \Gvac_{\mathrm{B}\Delta\mathrm{A}}
  \GG^{\mathrm{vac}\star}_{\Delta\mathrm{AB}}\vec{b}_{i}}$, which is
nonnegative by virtue of the positive-semidefiniteness of
$\Gvac_{\mathrm{B}\Delta\mathrm{A}}
\GG^{\mathrm{vac}\star}_{\Delta\mathrm{AB}}$. Therefore, any increase
in the volume of a body will increase the singular values of
$\Gvac_{\mathrm{BA}}$.

\section{Alternative bounds incorporating far-field radiative losses through
  $\Im(\Gvac_{\mathrm{AA}})$}

In this section, we derive an alternative bound
to~\eqref{Phiboundrhosingvals} that involves
$\Im(\Gvac_{\mathrm{AA}})$, thus capturing constraints on scattering
losses purely from finite object sizes rather than through multiple
scattering.  Starting from $\TT_{\mathrm{A}}^{-1} =
\VV_{\mathrm{A}}^{-1} - \Gvac_{\mathrm{AA}}$, operator manipulations
allow for writing $\TT_{\mathrm{A}}^{-1}
\Im(\VV_{\mathrm{A}}^{-1\star}) \TT_{\mathrm{A}}^{\star} =
\Im(\TT_{\mathrm{A}}) - \TT_{\mathrm{A}} \Im(\Gvac_{\mathrm{AA}})
\TT_{\mathrm{A}}^{\star}$. This allows for
rewriting~\eqref{exactPhiTAYB} as
\begin{equation*}
  \Phi = \frac{2}{\pi\zeta_{\mathrm{B}}}
  \tr\left[\GG_{\mathrm{AB}}^{\mathrm{vac}\star}
    \YY_{\mathrm{B}}^{\star} \YY_{\mathrm{B}} \Gvac_{\mathrm{BA}}
    \left(\Im(\TT_{\mathrm{A}}) - \TT_{\mathrm{A}}
    \Im(\Gvac_{\mathrm{AA}}) \TT_{\mathrm{A}}^{\star}\right)\right],
\end{equation*}
which now hides reciprocity, as no similar transformation has been
made to eliminate terms giving rise to $\zeta_{\mathrm{B}}$. Using the
singular value decomposition from~\eqref{opSVD} for $\YY_{\mathrm{B}}$
and $\Gvac_{\mathrm{BA}}$ but leaving $\TT_{\mathrm{A}}$ general, we
find that the constraint on $y_{i}$ is saturated when $y_{i} =
\left(\zeta_{\mathrm{B}}^{-1} + \zeta_{\mathrm{A}}^{-1} g_{i}^{2}
\bracket{\vec{a}_{i}, \TT_{\mathrm{A}} \TT_{\mathrm{A}}^{\star}
  \vec{a}_{i}}\right)^{-1}$. Completing the square, one may write
\begin{multline*}
  \Im(\TT_{\mathrm{A}}) - \TT_{\mathrm{A}} \Im(\Gvac_{\mathrm{AA}})
  \TT_{\mathrm{A}}^{\star} = \Im(\Gvac_{\mathrm{AA}})^{-1/2} \\ \times
  \Bigg[\frac{1}{4} \II_{\mathrm{A}} -
    \left(\Im(\Gvac_{\mathrm{AA}})^{1/2} \TT_{\mathrm{A}}
    \Im(\Gvac_{\mathrm{AA}})^{1/2} - \frac{\im}{2}
    \II_{\mathrm{A}}\right) \times
    \\ \left(\Im(\Gvac_{\mathrm{AA}})^{1/2} \TT_{\mathrm{A}}
    \Im(\Gvac_{\mathrm{AA}})^{1/2} - \frac{\im}{2}
    \II_{\mathrm{A}}\right)^{\star}\Bigg]\Im(\Gvac_{\mathrm{AA}})^{-1/2}.
\end{multline*}
This strongly suggests that the optimal $\TT_{\mathrm{A}}$ should be
diagonalized in the same basis as $\Im(\Gvac_{\mathrm{AA}})$, so if we
write
\begin{equation*}
  \Im(\Gvac_{\mathrm{AA}}) = \sum_{i} \rho_{i} \ket{\vec{q}_{i}}
  \bra{\vec{q}_{i}},
\end{equation*}
then one may also write $\TT_{\mathrm{A}} = \im\sum_{i} \tau_{i}
\ket{\vec{q}_{i}} \bra{\vec{q}_{i}}$. This implies that the constraint
on the singular values of $\YY_{\mathrm{B}}$ becomes
\begin{equation*}
y_{i} = \left(\zeta_{\mathrm{B}}^{-1} + \zeta_{\mathrm{A}}^{-1}
g_{i}^{2} \sum_{j} \tau_{j}^{2} |\bracket{\vec{a}_{i},
  \vec{q}_{j}}|^{2}\right)^{-1}
\end{equation*}
so $y_{i}$ depends on $\tau_{j}$ for every channel $j$, not just $j =
i$. Consequently, we arrive at the following bound:
\begin{equation}
  \Phi = \frac{2}{\pi} \sum_{i} \frac{\zeta_{\mathrm{B}} g_{i}^{2}
    \left(\sum_j \tau_{j} (1 - \rho_{j} \tau_{j}) |\langle \vec{a}_i,
    \vec{q}_j\rangle|^2\right)}{\left[1 + \zeta_{\mathrm{A}}^{-1}
      \zeta_{\mathrm{B}} g_{i}^{2} \left(\sum_{j} \tau_{j}^{2}
      |\bracket{\vec{a}_{i}, \vec{q}_{j}}|^{2}\right)\right]^{2}}
\end{equation}
for which finding the optimal values of $\tau_{i}$ for each channel
$i$ requires self-consistently solving a large set of nonlinear
equations subject to the constraint $\tau_{i} \leq \zeta_{\mathrm{A}}$
for each $i$. While this expression should yield tighter bounds on RHT
owing to the incorporation of constraints on scattering losses for
both objects in isolation (in addition to multiple scattering), it
appears to be analytically intractable and must therefore be evaluated
numerically, which we leave to future work; that said, numerical
solution of the optimal values of $\tau_{i}$ for evaluating this bound
requires solving a large set of nonlinear polynomial equations, which
is generally computationally easier than brute-force optimization of
the RHT spectrum due to the non-polynomial dependence of $\Phi$ on the
T-operators in general. We point out that in the nonretarded
quasistatic limit, $\Im(\Gvac_{\mathrm{AA}}) \to 0$, so all of its
singular values may be taken to vanish as well; this means that its
singular vectors become arbitrary, allowing for choosing
$\ket{\vec{q}_{i}} = \ket{\vec{a}_{i}}$. Doing so, the above
expression simplifies to $\Phi = \frac{2}{\pi} \sum_{i}
\frac{\zeta_{\mathrm{B}} g_{i}^{2} \tau_{i}}{(1 +
  \zeta_{\mathrm{A}}^{-1} \zeta_{\mathrm{B}} g_{i}^{2}
  \tau_{i}^2)^{2}}$, from which it can be seen that if the material
bound $\tau_{i} \leq \zeta_{\mathrm{A}}$ is saturated, each
contribution is identical to the corresponding contribution
from~\eqref{Phiboundrhosingvals}; if the material bound is not
saturated, then the optimal $\tau_{i} =
\frac{1}{\sqrt{3\zeta_{\mathrm{A}}^{-1} \zeta_{\mathrm{B}}} g_{i}}$
leads to a larger contribution than what we find
in~\eqref{Phiboundrhosingvals}, yielding an overall looser bound. This
corroborates the notion that our present bounds, which do not start
with an explicit expression in terms of $\Im(\Gvac_{\mathrm{AA}})$, do
not fully account for dissipation via far-field radiative losses,
while consideration of the singular values $\rho_{i}$ will remedy
this.

\nocite{apsrev41Control} \bibliographystyle{apsrev4-1}

\bibliography{generalheatboundspaper}
\end{document}